\documentclass[a4paper,11pt]{article}
\pdfoutput=1 

\usepackage{jcappub} 
\usepackage{siunitx}

\title{QUBIC V: Cryogenic system design and performance}

\author[1,2]{S.~Masi}
\author[1,2]{E.S.~Battistelli}
\author[1,2]{P.~de~Bernardis}
\author[3]{C.~Chapron}
\author[1,2]{F.~Columbro}
\author[1,2]{A.~Coppolecchia}
\author[1,2]{G.~D'Alessandro}
\author[1,2]{M.~De~Petris}
\author[3]{L.~Grandsire}
\author[3]{J.-Ch.~Hamilton}
% ...

\affiliation[1]{Universit\a`{a} di Roma - La Sapienza, Roma, Italy}
\affiliation[2]{INFN sezione di Roma, 00185 Roma, Italy}
\affiliation[3]{Universit\'e de Paris, CNRS, Astroparticule et Cosmologie, F-75006 Paris, France}
% ...

\abstract{
Current experiments aimed at measuring the polarization of the Cosmic Microwave Background (CMB) use cryogenic detector arrays with cold optical systems to boost their mapping speed. For this reason, large volume cryogenic systems with large optical windows,  working continuously for years, are needed. The cryogenic system of the QUBIC (Q \& U Bolometric Interferometer for Cosmology) experiment solves a combination of simultaneous requirements: very large optical throughput ($\sim$40~cm$^2$sr), large volume ($\sim$1~m$^3$) and large mass ($\sim$165~kg) of the cryogenic instrument. Here we describe its design, fabrication, experimental optimization and validation in the Technological Demonstrator configuration. The QUBIC cryogenic system is based on a large volume cryostat that uses two pulse-tube refrigerators to cool the instrument to $\sim$3~K. The instrument includes the cryogenic polarization modulator, the corrugated feedhorn array, and the lower temperature stages: a $^4$He evaporator cooling the interferometer beam combiner to $\sim$1~K and a $^3$He evaporator cooling the focal-plane detector arrays to $\sim$0.3~K. The cryogenic system has been tested and validated for more than 6 months of continuous operation. The detector arrays have reached a stable operating temperature of 0.33~K, while the polarization modulator has operated at a $\sim$10~K base temperature. The system has been tilted to cover the boresight elevation range 20$^\circ$-90$^\circ$ without significant temperature variations. The instrument is now ready for deployment to the high Argentinean Andes.
}


\arxivnumber{2008.10659}

\begin{document}
\maketitle
\flushbottom

\section{Introduction}
\label{sec:intro}
Cosmic Microwave Background (CMB) polarization is a very important tool for investigating several phases of the evolution of the universe, including the very early inflation phase (see e.g. \cite{2016ARA&A..54..227K}). For this reason, the search for CMB polarization is vigorously pursued by many experiments worldwide. The expected signal is very small, in the sub $\mu K_{rms}$ range, and large angular scales are involved. For these reasons CMB polarimeters must feature high sensitivity and, once the sensitivity is limited by photon noise, high mapping speed, combined with extremely demanding accuracy and stability. Most of the current experiments are scanning imagers (see e.g. \cite{2018JLTP..193.1112G, 2020arXiv200501187L, 2020JLTP..tmp...95S, 2012JLTP..167..859B} ). The Q \& U Bolometric Interferometer for Cosmology (QUBIC) \citep{2020JLTP..tmp...35S, 2019Univ....5...42M, 2020AA.QUBIC.PAPER1}, is a bolometric interferometer, combining the high mapping speed of bolometer arrays with the beam purity and control of systematic effects of interferometers. A block-diagram of the instrument is shown in the top panel of figure \ref{fig:optisketch}. The main characteristics are reported in table \ref{table:qubic_params}. 
In order to achieve long integrations and minimize atmospheric emission and noise, the instrument will operate from a site in the high Argentinean Andes, at an altitude of $\sim$ 5000m, exploiting its thin and cold atmosphere.

As all modern CMB experiments using bolometer arrays, the detectors of QUBIC require a large sub-K cryogenic system to operate at the required level of sensitivity. The development of custom cryogenics for CMB receivers has a long history. The need of fast sky mapping, involving large fields of view and collecting areas, drove the development of increasingly large cryogenic systems (see e.g. \cite{1999Cryo...39..217M, 2009JPhCS.155a2008D, 2011AeA...536A...2P, 2015Cryo...72...65G, 2018SPIE10708E..1HS, 2020Cryo..11003129C, 2020arXiv200909997D}). 

Sub-K cooling can be obtained by means of simple sealed L$^4$He and L$^3$He evaporation refrigerators (EVRs, see e.g. \cite{1985RScI...56..318T, 1998Cryo...38..319M, 2018SPIE10708E..3VM}), if an operating temperature $\sim$300~mK is sufficient. The detectors of QUBIC operate in two bands, centered at 150 and 220~GHz, around the maximum brightness of the CMB. The bandpasses (130-170~GHz and 200-300~GHz) match the width of the atmospheric windows, exploiting the low emissivity (typically $\sim 2\%$ and $\sim 4\%$, respectively) of the cold ($\sim$240~K), rarefied atmosphere above the site. The integrated radiative load from the atmosphere, for single-mode detectors, is of the order of 2 and 10~pW at 150 and 220~GHz, respectively (assuming 40\% global efficiency), and the photon noise (neglecting turbulence) is of the order of $2\times10^{-17}~W/\sqrt{Hz}$ and $7\times10^{-17}~W/\sqrt{Hz}$ respectively. Similar additional loads (and photon noises) are expected from the optical window and the filter chain. This means that detectors with NEP $\sim$$10^{-17}~W/\sqrt{Hz}$ can be used to obtain photon-noise-limited performance. Such an NEP can be achieved using TES detectors operating at 0.3~K (see \cite{2020AA.QUBIC.PAPER4}). Detector operation at lower temperatures would not significantly improve the overall performance of the instrument. This means that we can avoid the complexity of dilution and adiabatic demagnetization refrigerators, improving the reliability of the instrument. The $^3$He refrigerator is used to cool the detector arrays, while the $^4$He refrigerator is used to cool the optical beam combiner and the filter stack, reducing their radiative loading on the detectors. Due to the large optical throughput of the system, the volume of the 1-K section of the instrument is of the order of 1~m$^3$. 

A pre-cooling stage, with a temperature of $\sim$3~K, and with sufficient cooling power, is required to condense $^4$He and $^3$He in the evaporation refrigerators and start their cooling cycle. In this phase the precooler must withstand the large ($\sim$0.5~W) power dissipated to heat the cryopumps and desorb the gas. Moreover, the array of 400 feed-horns (which represent the apertures of the interferometer) and the half wave plate (HWP) polarization modulation unit must be cooled to a similar temperature to reduce their radiative loading on the bolometers. The only viable solution to ensure long-duration, continuous, unmanned cryogenic operation at $\sim$3~K, in a remote high altitude environment where human operations are challenging and delivery of liquid cryogens is difficult, makes use of mechanical cryocoolers, so that the cryogenic operations can be controlled and monitored remotely. Pulse Tube (PT) cryocoolers feature reduced vibration levels compared to Gifford McMahon cryocoolers (see e.g. \cite{PTvsGM}). The use of PTs for CMB experiments located in remote sites started a long time ago (see e.g. the BRAIN instrument in Dome-C (Antarctica) \cite{2007NewAR..51..256P}), and is now standard (see e.g. \cite{2010SPIE.7741E..1RS, 2011PASP..123..568C}). QUBIC uses two two-stage PTs, each delivering $\sim$0.9~W of cooling power at $\sim$4~K and $\sim$35~W at $\sim$45~K. The coldest stage ($\sim$3~K) is used as described above, while the second stage is used mainly to reduce the radiative load from the large surface (a few m$^2$) shield surrounding the 3-K stage. In addition, it cools the optical filters limiting the radiative load entering the large diameter ($\sim$0.5~m) optical window. Moreover, it provides an intermediate heat sink for the electrical harness required to read the detector arrays. 

Due to the large mass of the instrument section to be cooled to 1~K or below ($\sim$165~kg), careful control of the heat flows is mandatory. For this reason heat switches and high conductivity heat paths are very important parts of this cryogenic system.

In this paper, which is part of a set describing the current status of the QUBIC experiment \cite{2020AA.QUBIC.PAPER1, 2020AA.QUBIC.PAPER2, 2020AA.QUBIC.PAPER3, 
2020AA.QUBIC.PAPER4, 2020AA.QUBIC.PAPER6,
2020AA.QUBIC.PAPER7, 2020AA.QUBIC.PAPER8}
 we describe the design, fabrication, experimental optimization and validation of the QUBIC cryogenic system in the Technological Demonstrator (TD) configuration. This configuration differs from the Full Instrument (FI) configuration because the optical apertures are reduced in size, only one of the two focal planes (the transmitted one) is installed, and only 1/4 of the detectors (one wafer) is present. The cryogenic system, however, is already the final one. 

\section{Main cryostat}

\subsection{General design}

\begin{figure}
  \centering
  \includegraphics[width=0.7\linewidth]{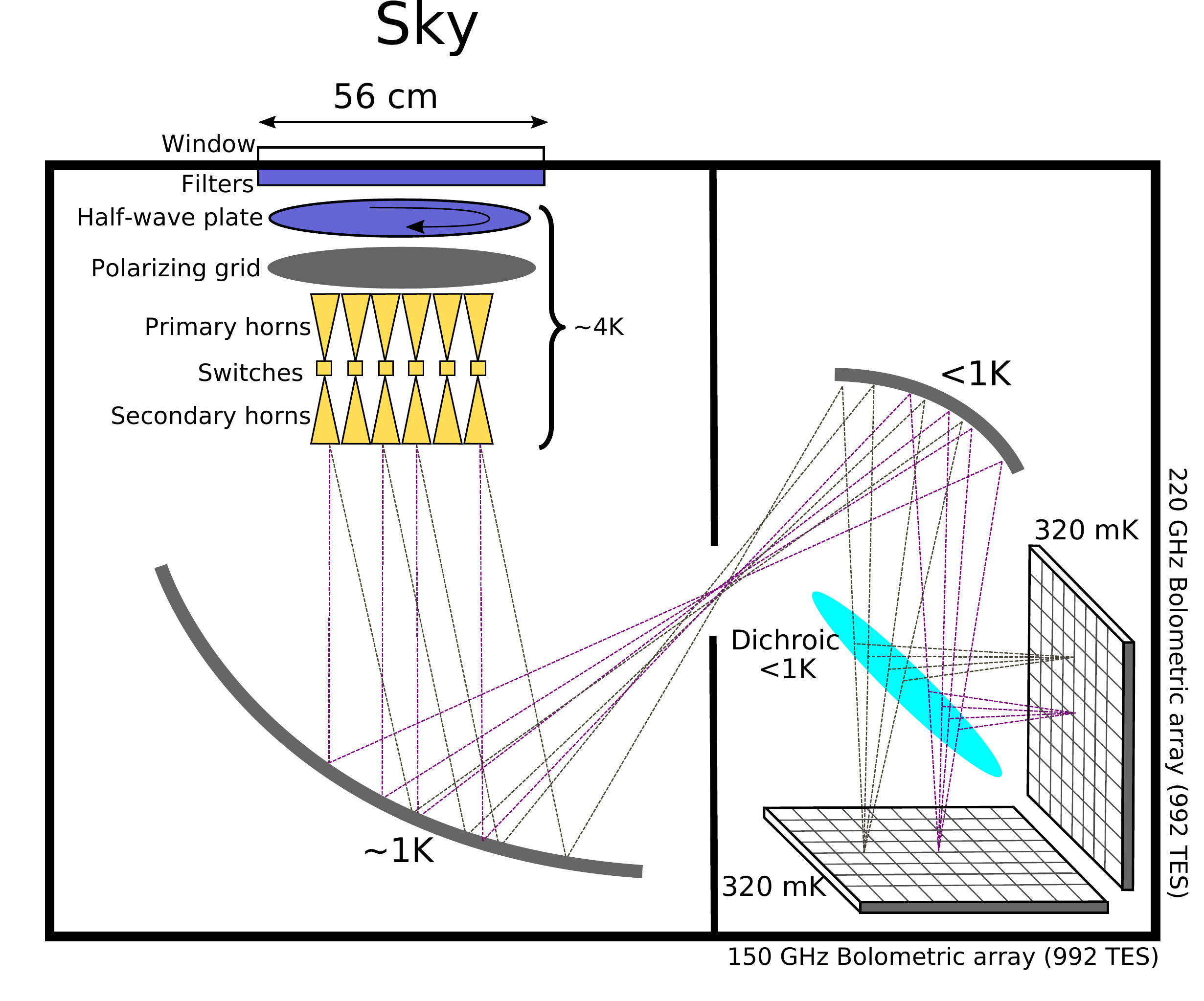}
  \includegraphics[width=1\linewidth]{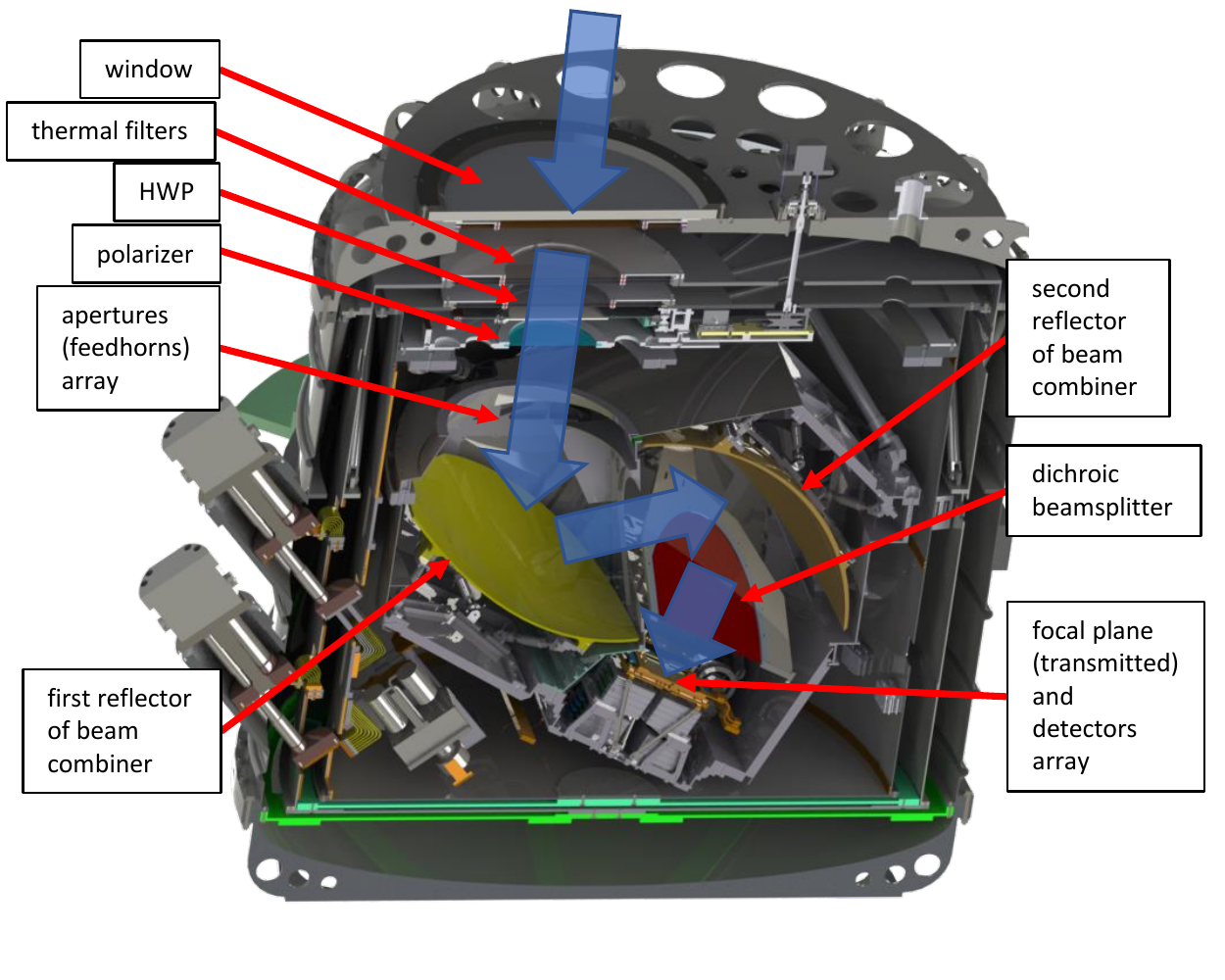}
  \caption{{{\bf Top:} Schematic of the QUBIC instrument. {\bf Bottom:} Cut-view of the QUBIC cryostat. The wide arrows indicate the optical path. The main optical components are indicated by the labels. \label{fig:optisketch}}
  }
\end{figure}

A cut view of the QUBIC instrument cryostat, displaying the location of the main components inside the cryostat, is shown in the bottom panel of figure \ref{fig:optisketch}.  

\newlength{\colone}
\settowidth{\colone}{Synthesized beam FWHM [degrees]~~~}
\newlength{\coltwo}
\settowidth{\coltwo}{[131-169]~GHz}
\newlength{\colthree}
\settowidth{\colthree}{0.39 (150~GHz), 0.27 (220~GHz)}

\begin{table}[t]
    \renewcommand{\arraystretch}{1.}
    \begin{center}
        \caption{\label{table:qubic_params}  Main parameters of QUBIC in the technology demonstrator version (QUBIC-TD) and in the full instrument version (QUBIC-FI).}
\begin{tabular}{p{\colone} p{\coltwo} p{5.5cm}}
          \hline
            Parameter& QUBIC-TD & QUBIC-FI \\
            \hline
            \hline
            Frequency channels \dotfill &150 GHz & 150 GHz \& 220 GHz\\
            Frequency range 150 GHz \dotfill &[131-169] GHz &[131-169] GHz\\
            Frequency range 220 GHz \dotfill &- &[192.5-247.5] GHz\\
            Window Aperture [m]\dotfill & 0.56 & 0.56 \\
            Number of horns\dotfill &64 &400\\
            Number of detectors\dotfill &248 &992$\times$2\\
            Detector noise [$\mathrm{W/\sqrt{Hz}}$]\dotfill & 2.05$\times 10^{-16}$ & 4.7$\times 10^{-17}$ \\
            Focal plane temp. [mK]\dotfill &300 &300\\
            Sky Coverage\dotfill &1.5\% &1.5\%\\
            Synthesized beam FWHM [degrees]\dotfill &0.68 &0.39 (150~GHz), 0.27 (220~GHz)\\
        \end{tabular}
        
    \end{center}
\end{table}

Following the optical path of the measured radiation, the system is composed of
\begin{itemize}
\setlength\itemsep{-0.3em}
\item a large (0.56-m diameter) optical window
\item quasi-optical filters rejecting radiation outside the bands of interest
\item a HWP (with its rotation system) and a large-diameter wire-grid polarizer, constituting a Stokes polarimeter 
\item an array of corrugated feedhorns, representing the apertures of a compact aperture-synthesis interferometer, complemented by individual opto-mechanical switches, useful for self-calibration
\item a two-mirror, large-throughput beam combiner 
\item a tilted dichroic, reflecting and transmitting the beams towards two separate focal planes operating at different frequencies
\item two focal planes, populated with bolometric detector arrays, where the optical fringes are formed and detected, each with its own band-defining filter. 
\end{itemize}

\noindent The most relevant features, from the cryogenic point of view, are:
\begin{itemize}
\setlength\itemsep{-0.3em}
\item The continuous operation requirement, at elevation angles between 20$^\circ$ and 80$^\circ$, in an unmanned harsh environment.
\item The large volume ($\sim$1~m$^3$) and mass ($\sim$165~kg) of the instrument to be cooled to $\sim$1~K, driving the overall dimensions of the cryostat.
\item The large area of the optical window and the large throughput of the beam entering the window and reaching the detectors at 0.3~K, requiring large quasi-optical filters to reject the radiative load entering the window. 
\item The low temperature ($<$4~K) of the condensers for $^3$He and $^4$He in the evaporation refrigerators to be attained for efficient condensation while the cryopumps are at $>$30~K. 
\item The presence of a polarization modulator, cooled by the 3-K stage to reduce its emission, and producing heat load pulses from friction during its intermittent and stepped rotations. 
\item The presence of a large number of detectors and sensors, requiring a complex harness for the readout electronics, and a large number of vacuum feedthroughs. Moreover, a careful reduction of vibrations is needed to avoid microphonic noise in the detectors.
\end{itemize}
Our design implements original solutions to cope with the requirements listed above. In particular, we use two PT refrigerators, one taking care of the large radiative load on the 40-K and 3-K stages, and the other one dedicated to the sub-K coolers. We use a sequence of carefully selected filters to limit the radiative load through the optical path. Moreover, we use an original vibration-damping system for the rotary valves of the PTs, and a set of advanced, optimized heat switches to operate the system.

\subsection{Mechanical design}

\subsubsection{Vacuum shell}

The purpose of the outer shell of the cryostat is to maintain the evacuation of the internal volume, so that there is no heat exchange, mediated by gas, between the room temperature environment and the cryogenic instrument. This means that the outer shell must withstand a pressure difference of $\sim$100~kPa, and be perfectly leak tight. Given the large instrument volume ($\sim$1~m$^3$), the mechanical loads on the shell are very large. For example, the load on the $\sim$1.4~m  diameter top flange is $\sim$150~kN. For transportation and orientation purposes, the mass of the cryostat should be minimized. We opted for an aluminum alloy (Al5083) for construction.

\begin{figure}
  \centering
  \includegraphics[width=0.64\linewidth]{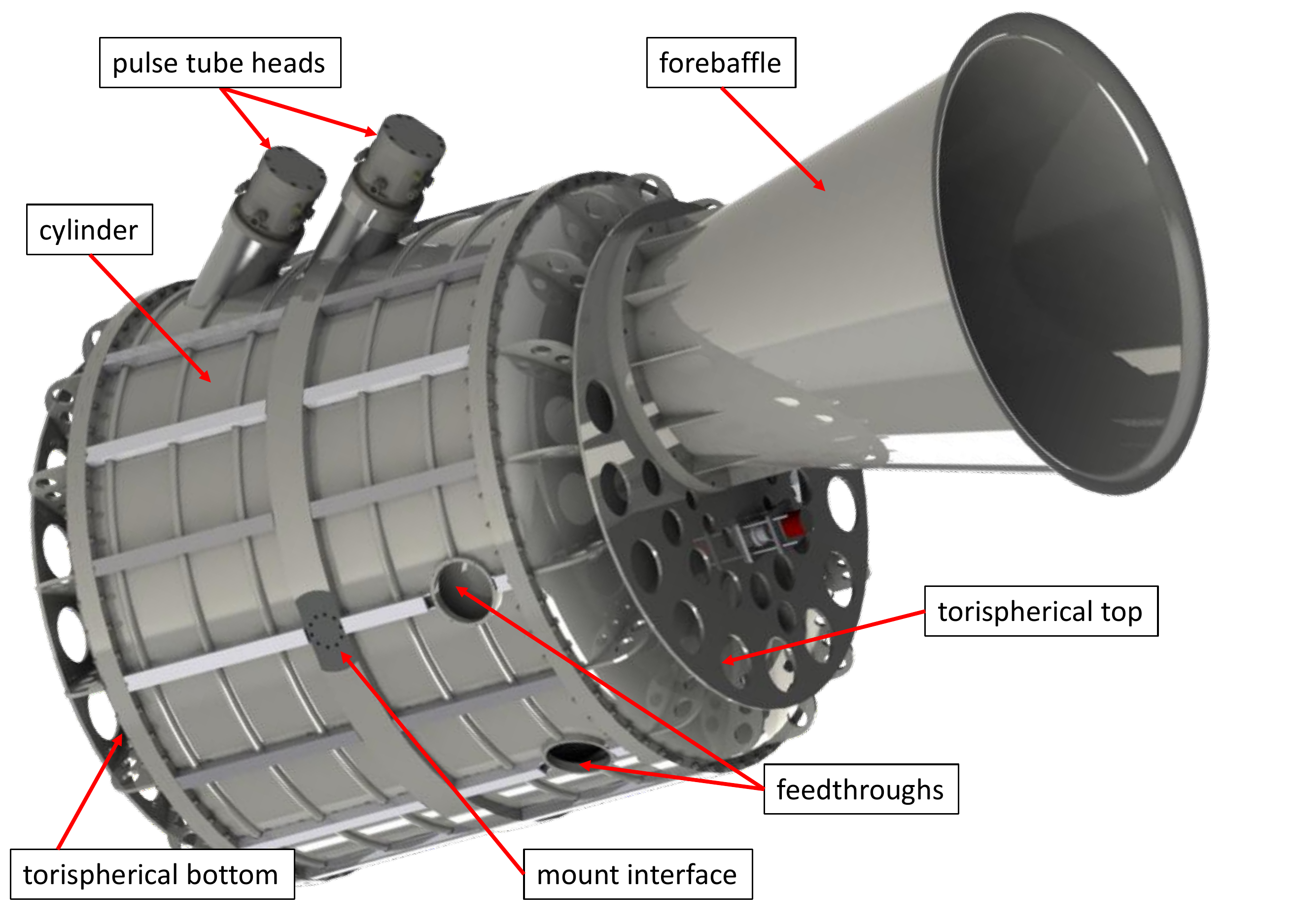} \includegraphics[width=0.35\linewidth]{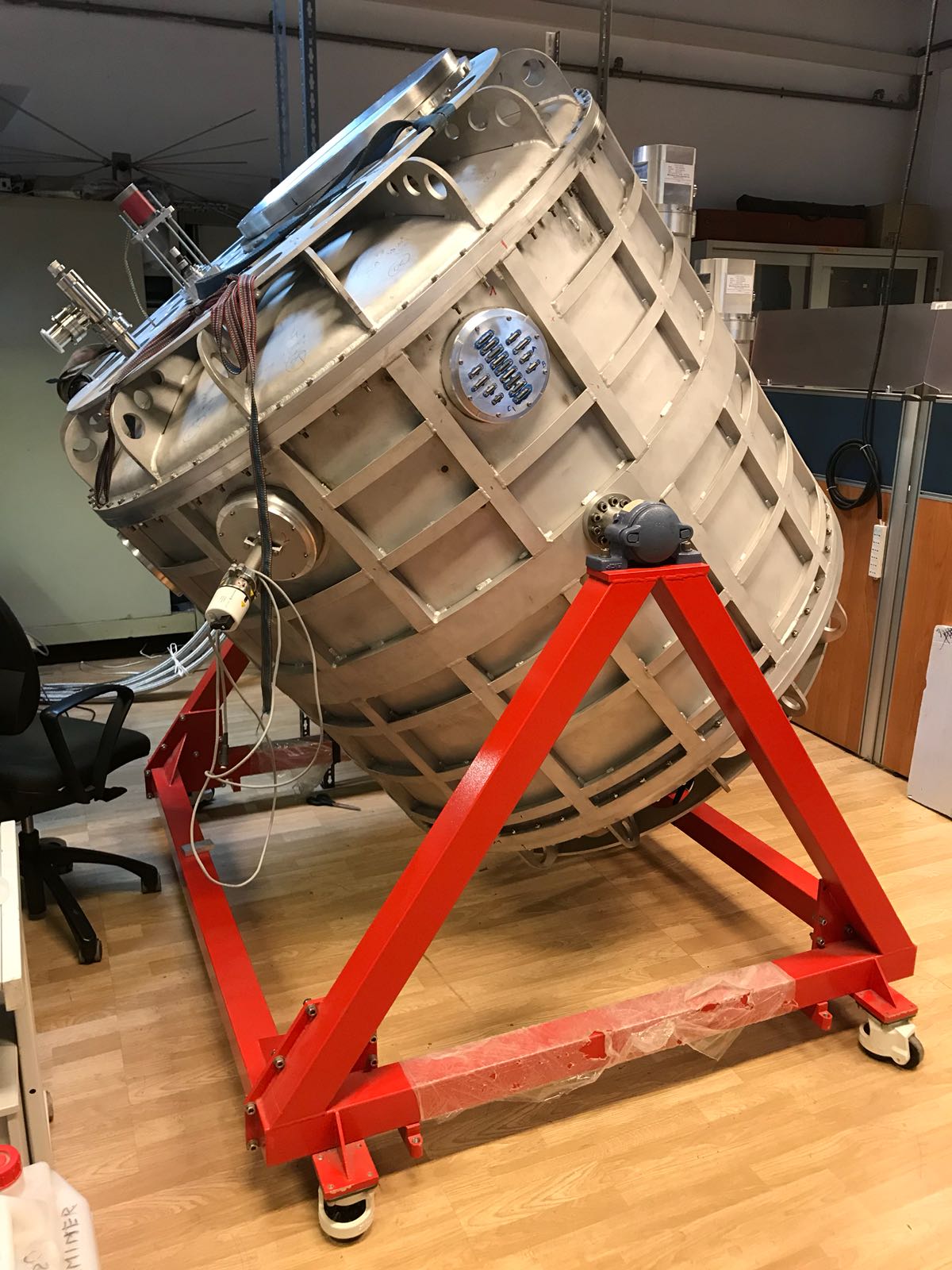}
  \caption{{\bf Left:} Sketch of the outer (vacuum) shell of the QUBIC cryostat, approximately 1.4~m in diameter and 1.6~m tall (forebaffle excluded). The parts are made of Al5083 aluminum alloy, welded together. The lightweight design has been optimized by extensive use of finite-element simulations. {\bf Right:} Picture of the assembled cryostat, without the forebaffle. \label{fig:shell}}
\end{figure}

The outer shell is a cylinder (1420~mm in diameter, 1112-mm tall), with torispherical top and bottom ends. We minimized the solid thickness of the torispherical head by using a thin (5~mm) torispherical sheet welded to a lightweight, flat, 10-mm thick disk, through 16 radial vertical ribs (see figure \ref{fig:shell}). The top flange accommodates a 600-mm diameter aperture for the optical window. 

Vacuum sealing between the 25-mm thick UHMW-PE \cite{2018InPhT..90...59D} optical window and the top flange is guaranteed by a Buna-N O-ring, with 5-mm diameter cross-section.  In addition, for optional operation at extremely low temperatures, indium grooves are present. The same type of O-ring and indium grooves is used to obtain the vacuum seal between the torispherical flanges and the cylinder. Additional flanges are present for the vacuum valve, and the HWP mechanism magnetic joint \cite{2020AA.QUBIC.PAPER6}. 

The cylinder is obtained from a roll-bent 3.5-mm thick sheet, reinforced by welded vertical and longitudinal ribs. The cylinder includes two slanted lateral flanges, tilted by 40$^\circ$ from the vertical, on to which to mount the two PT heads (see figure \ref{fig:shell}). Their tilt has been chosen so that they operate vertically in the middle of the elevation range of the cryostat. In this way the tilt of the two PTs with respect to the vertical does not exceed 30$^\circ$ in the entire elevation range of interest. Roughly in the middle of the cylinder there is a thick belt with two opposite flat flanges and tap holes circles (8 $\times$ M10 on a 100-mm diameter) to interface the cryostat to the mount. In addition, five 120-mm diameter apertures are present on the cylinder for the electrical feedthrough flanges (see figure \ref{fig:shell}). The thickness of the sheets, the number of ribs, and their size have been optimized through extensive finite-element simulations. 

The total mass of the optimized outer shell is $\sim$225~kg, representing $\sim$50$\%$ of the total mass of the cryostat. 

\subsubsection{Minimization of induced vibrations}\label{par:vibrations}

One potential drawback of the use of PT refrigerators is the presence of vibrations, synchronous with the periodic ($f\sim$1.7~Hz) aperture of the rotary valve and the consequent pressure wave propagating in the pulsed tubes. This results in small periodic variations of the position of the cold end (on the order of 10~$\mu$m p-p) and small periodic movements of the rotary valve assembly and the He lines. Moreover, vibrations from the compressor can propagate all the way to the cryostat via the flexible He gas lines. 

Vibrations induce microphonic noise in two ways. Shaking the detector wafer, the signal wires and connectors, and the proximity electronics, produces noise caused by acceleration of the signal carriers, by modulation of capacitive couplings to the ground and by triboelectricity effects. Moving (decentering) optical apertures modulates the radiative background on the detectors. Since the background is much larger than the signals to be detected, very small decentering can be important, especially where the edge taper is not very high. 

\begin{figure}
  \centering
  \includegraphics[width=0.55\linewidth]{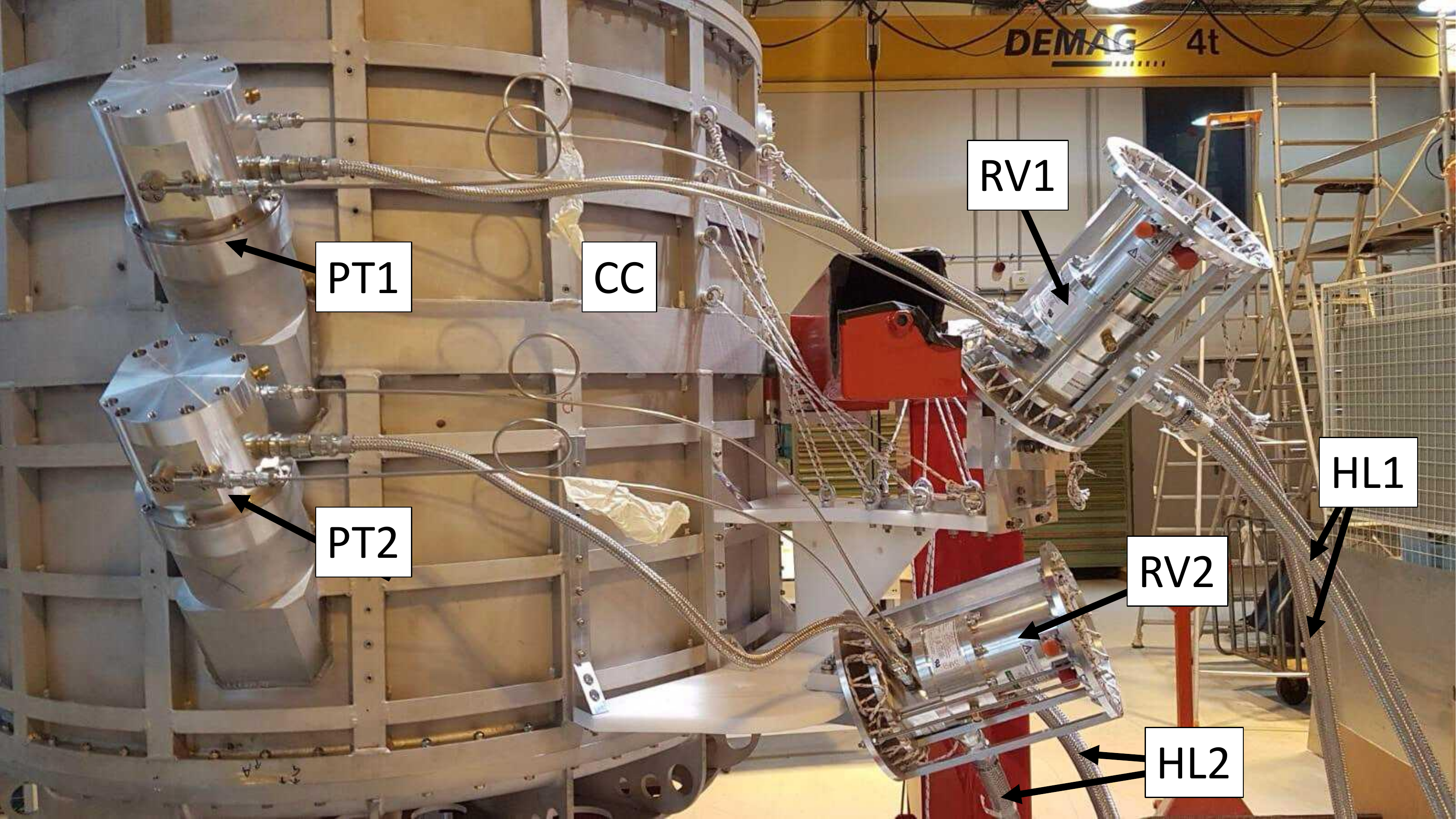}
  \includegraphics[width=0.40\linewidth]{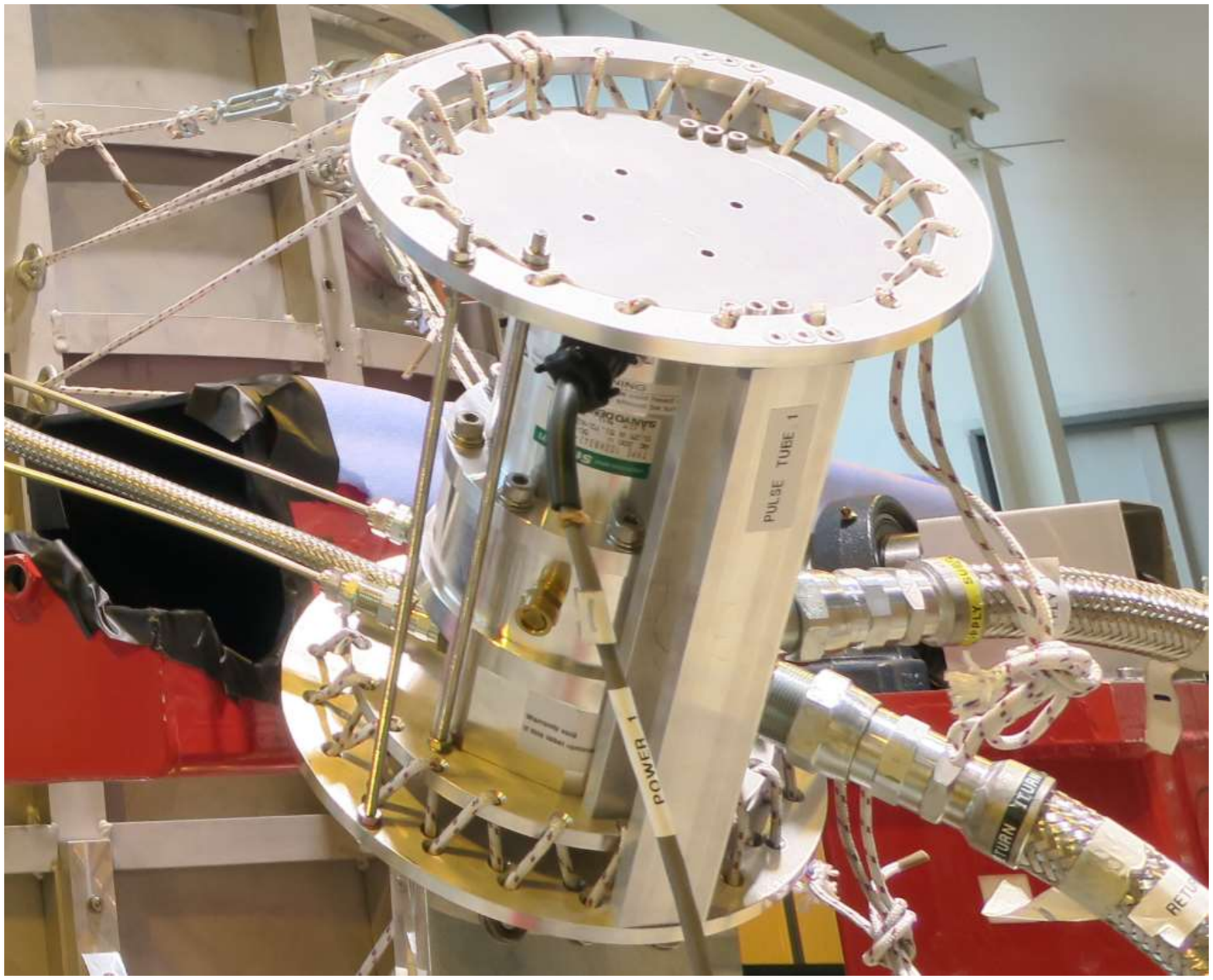}
  \caption{{\bf Left:} Picture of the rotary valves RV1 and RV2 of the pulse tubes PT1 and PT2. Also shown is their suspension system, based on nylon cords to dissipate the mechanical vibration energy. The He flexlines HL1 and HL2 on the right go to the PT compressors without touching the cryostat shell CC. {\bf Right:} Zoomed-in view of a rotary valve assembly with its suspension system. \label{fig:rotaryvalve}}
\end{figure}

In our implementation we mitigate the transmission of vibrations in two ways. The rotary valves are separated from the cryostat body and supported by a system of nylon cords, dissipating the mechanical vibration energy, as seen in figure \ref{fig:rotaryvalve}. Moreover, the He gas lines from the PT compressor to the rotary valve do not touch the cryostat body.

In addition, the 40-K and 3-K cold heads have been decoupled from the 40-K and 3-K stages by means of flexible heat straps, as detailed in \S \ref{par:40K3Kstages}. The flexible straps have also the important purpose of compensating differential thermal contractions between the cold stages and the outer shell, which otherwise would produce strong mechanical loads on the tubes of the PTs. The effectiveness of this solution is discussed in a companion paper \cite{2020AA.QUBIC.PAPER4}.
\subsubsection{40-K and 3-K stages and supports}\label{par:40K3Kstages}
\begin{figure}
\centering
\includegraphics[width=1\linewidth]{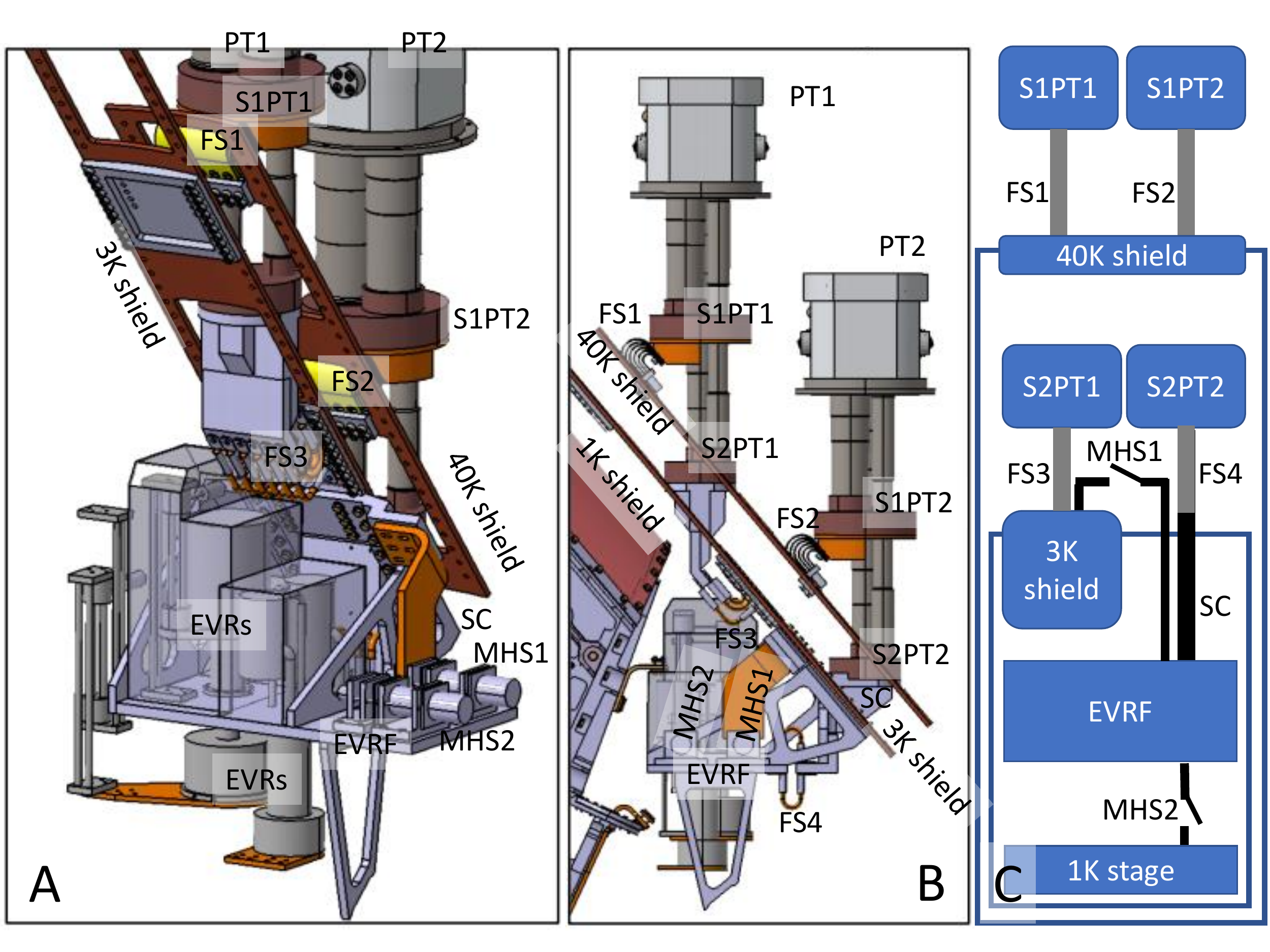}
\caption{Sketches (panels A and B) and block diagram (panel C) of the connections between the two pulse tubes and the 40-K, 3-K, 1-K stages. Both PTs cool the 40-K shield through the flexible copper sheet assemblies FS1 and FS2, connecting the 40-K shield to their first stages S1PT1 and S1PT2. The second stage of PT1 (S2PT1) cools the 3-K shield by means of the flexible copper braids FS3. The second stage of PT2 (S2PT2) cools the evaporation refrigerators (EVRs) flange (EVRF) through a solid connection (SC) and a set of flexible copper braids (FS4), connected to the EVRF. A mechanical heat switch (MHS1) can be closed to to make the PT second stages (S2PT1 and S2PT2) operate in parallel for a larger cooling capacity on the EVRF, when needed (typically when the EVR cryopumps are heated up to desorbe gas, and the temperature of the EVRF must remain low to allow for an efficient condensation). A second mechanical heat switch (MHS2) can be closed to connect directly the 1-K stage to the 3-K stage (EVRF) to speed up its initial cooldown to 3~K. MHS2 will be open afterwards to allow further cooling through the 1-K evaporation refrigerator. The thermal architecture of the 1-K and 0.3-K stages is summarized below, in figure \ref{fig:subKarchitecture}. } \label{fig:PTs} 
\end{figure}

The 40-K stage is a shield cooled by the first stage of the PTs, and has the important purpose of reducing the radiative load from the 300-K shell, while mechanically supporting the 3-K stage. The shield itself is a 3-mm thick, roll-bent Al5083 cylinder, 1220~mm in diameter and 1088-mm tall. The part of the cylinder closer to the pulse-tube stages is sandwiched between two vertical copper flanges, as tall as the entire cylinder, by means of a large number (66) of M5 bolts and nuts (see figure \ref{fig:PTs}). The vertical copper flange is thermally connected to the first stages of both PTs. The purpose of this copper section is to collect the heat flow from the entire shield and reduce thermal gradients across the shield. The top part of the 40-K stage is supported by a set of 16 tubular struts, made in G10 fiberglass for low heat conductivity. These struts run from 300~K to 40~K, and are 350-mm long with 24~mm outer diameter and 20~mm inner diameter. Each tube has two aluminum cores at the ends, glued by means of epoxy resin. We have tested the tensile strength of these joints in liquid nitrogen, using up to 2500~N of tension. The bottom part of the stage is supported by a set of 8 radial G10 fiberglass straps ($\sim$0.5-m long, 6~mm~$\times$~40~mm cross-section). 
\begin{figure}
\centering
\includegraphics[width=0.9\linewidth]{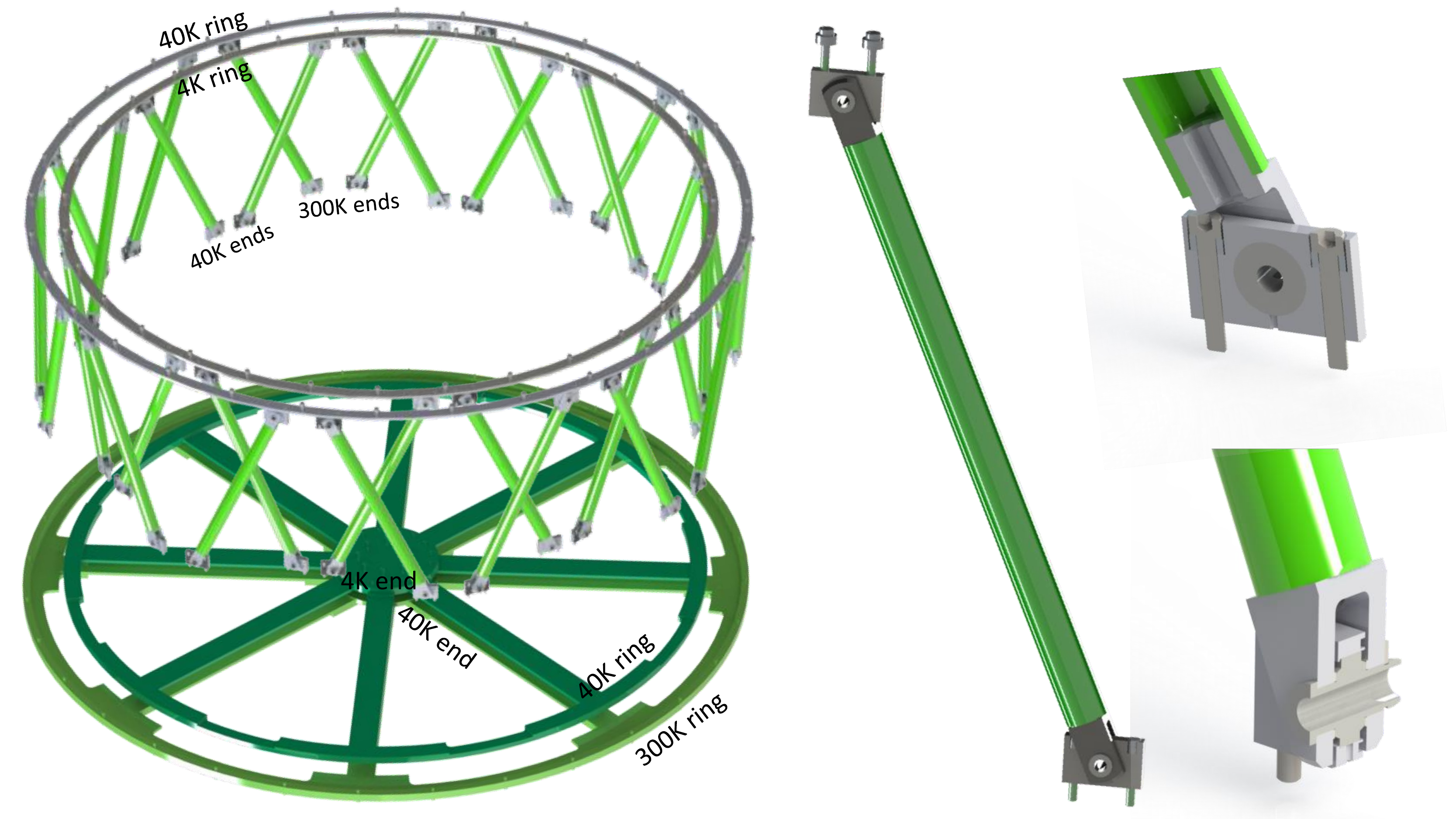}
\caption{{\bf Left:} Sketch of the drum-like structures supporting the top part of the 40-K and 3-K stages of the instrument (top) and of the straps structures supporting the bottom part of the 40-K and 3-K stages (bottom). Each drum-like structure is obtained by assembling 16 tubes in G10 fiberglass. {\bf Right:} Details of the tube joints, where the G10 tubes are glued to aluminum core ends with epoxy resin. \label{fig:drum}}
\end{figure}
These support structures are very stiff. For a lateral load of 1500~N on the 40-K stage cylinder we measured a lateral displacement of 0.5~mm, while for an axial load of 1500~N we measured an axial displacement of 0.3~mm.  

The 3-K stage is a shield cooled by the second stage of the PTs. It reduces the radiative load from the 40-K shell on the 1-K stage and provides the base temperature for condensation of $^4$He and $^3$He in the evaporation refrigerators. The shield is a 3-mm thick, roll-bent Al5083 cylinder, 1096~mm in diameter and 1074-mm tall. A vertical copper flange, similar in shape and purpose to the one used in the 40-K shield is present here as well. The 3-K stage is supported by the 40-K stage with the same type of G10 struts and straps supporting the 40-K stage. 

The 40-K and 4-K stages are connected to the first and second stages of the PTs via sets of flexible OFHC copper sheets and braids, as sketched in figure \ref{fig:PTs}. In particular, the flexible connection between the first stage of each PT and the 40-K shield is made using 6 blades of 1-mm thick OFHC copper, 50 to 120-mm long and 75-mm wide, achieving a thermal conductance of $\sim$6~W/K at 40~K while maintaining a flexible connection between the PT flange and the 40-K shield. The flexible connection between the second stage of each PT and the 3-K stages is made using 18(20) braids of OFHC copper, $\sim$50-mm long and 6~mm in diameter, achieving a thermal conductance of 2.7(3.0)~W/K at 4~K for PT1(PT2), while maintaining a flexible connection between the PT flanges and the 3-K stage. Each copper braid end is clamped using one M5 bolt.

\subsection{Thermal design}

The main cryostat is based on two Sumitomo SRP-082B2S-F70H pulse tube coolers. Each provides $\sim$0.9~W of cooling power at the second stage (at 4.2~K), and $\sim$35~W of cooling power at the first stage (at 45~K). 

\subsubsection{Loads on the 40-K stage}

The heat load on the first stage of the PT is dominated by radiation, due to its large area ($\sim6$~m$^2$) and the presence of a large optical window. A blanket of 30 aluminized mylar layers (with a total thickness of $\sim$15~mm) has been wrapped around the entire 40-K shield surface (excluding the optical beam aperture, which is less that 10\% of the total), drastically reducing the radiative load (see table \ref{tab:loadsmain}). However, a residual unavoidable radiative load comes from the large optical window. In fact, the need for a large-throughput beam entering the cryostat window and propagating all the way to the detector arrays poses important challenges from a cryogenic point of view. When operating in the 300~K radiative background of the laboratory, the 0.56~m diameter window collects $\sim$110~W of radiative power, mostly in the thermal IR band, spread over a 2$\pi$~sr solid angle. Only a few $\mu$W out of 110~W are in the frequency band (130-170 GHz and 200-240 GHz) and in the throughput ($\sim$0.05~sr) of interest for the astrophysical measurements, and must be propagated all the way to the detectors. All the rest should be rejected. The strategy adopted for mitigating the radiative heat load entering the window is to use quasi-optical filters along the optical beam path, to sequentially reject lower and lower frequency components above the band of interest. The filters are designed to reject most of the unwanted radiation by reflection. However, a small fraction of the radiation will be absorbed, producing a heat load on the cooling stages. This must be sunk mainly towards the 40-K stage, cooling the filters by contact to minimize their thermal emission. The filter chain of the QUBIC cryostat is detailed in table \ref{tab:filters}.
\begin{table}[htbp]
\centering
\caption{\label{tab:filters} Blocking filters for QUBIC. The indicated diameter is the optically-constrained clear aperture (given for both the FI and the TD). The \emph{height} column gives the distance of the center of each element from the aperture array, measured along the optical path. $\epsilon$ is the emissivity, averaged over the observation band. }
\smallskip
\begin{tabular}{|r|l|r|r|r|r|}
\hline
stage & filter type & diameter (TD/FI) & height & $\epsilon$ & filter T \\
 & & (TD/FI)(mm) & (mm) & (\%) &  (K) \\
\hline
shell & UHMW-PE window & 560/560 & 480 & 3& $\sim$ 270 \\
shell & IR blocker 1  &278/435 & 460 & 1 & $\sim$ 270 \\
shell & IR blocker 2 &278/435 & 452 & 1 & $\sim$ 270 \\
shell & foam MLI &278/435 & 415 & $<$1 & $\sim$ 220 \\ 
\hline
40-K & IR blocker 3 &225/430 & 342 & 1 & $\sim$ 100 \\
40-K & IR blocker 4 &225/430 & 335 & 1 & $\sim$ 100 \\
40-K & IR blocker 5 &225/430 & 327 & 1 & $\sim$ 100 \\
\hline
3-K & IR blocker 6 &200/410 & 286 & 1 & $\sim$ 10 \\
3-K & 12 cm$^{-1}$ low-pass edge  &200/410 & 276 & 2 & $\sim$ 10 \\
3K & Half Wave Plate &180/370 & 238 & 2.5&  $\sim$ 10 \\
3-K & polarizer &170/360 & 184 & 2 & $\sim$ 10 \\
3-K & horns and switches (square) & 120/400 & 0 & 5 & $\sim$ 10 \\
\hline
\end{tabular}
\end{table}
The window is a 25-mm thick slab of ultra-high molecular weight polyethylene (UHMW-PE). The two surfaces will be treated with an anti-reflection coating layer  made from porous PTFE foam  \cite{10.1117/12.856919}. Its emissivity is in the few \% range in the 150~GHz band \cite{2018InPhT..90...59D}. This remains essentially constant at higher frequencies, all the way to $\sim$3~THz. The emissivity increases gradually in the thermal IR. IR blockers (thermal shaders) are capacitive metal meshes \cite{2006SPIE.6275E..0TT}, preferentially reflecting radiation around the peak of the 300~K blackbody ($\sim$30~THz), with negligible absorption. In a stack of highly reflective thermal shaders, the first one will receive the full power from the external environment, reflect away most of it, and remain basically at room temperature. The ones that follow, due to the reduced heat input and the exposure of the internal side to a colder environment, will have gradually decreasing temperatures, as happens for the layers of a multi-layer-insulation blanket. The very thin ($\sim$3~$\mu$m)  polypropylene support sheet and the size of the metal squares in the capacitive metal mesh ($\sim$~10$\mu m$ side) make these layers effectively transparent to mm-waves. The sequence of thermal shaders in QUBIC removes more than 99\% of the input heat load (see e.g. \cite{2015Cryo...72...65G} for measurements on a similar system). Their performance is improved with the use of a stack of foam slabs, which act in a similar way (see e.g. \cite{2013RScI...84k4502C}). We inserted ten layers of Plastazote HD-30, $\SI{3}{\milli\meter}$ thick and spaced by $\SI{3}{\milli\meter}$ behind the window, as in \cite{bicep3_foam}. We estimate that only 11~W from the 300-K laboratory environment are transferred to the 40-K stage of the QUBIC cryostat through the window. This will reduce to 5W on site. The thermal filters are followed by low-pass edge metal mesh filters \cite{1994ApOpt..33.6046P, 2006SPIE.6275E..0UA} at decreasing temperatures and with decreasing cut-off frequencies. Additional band-defining filters are placed on the detector arrays as described in \cite{2020AA.QUBIC.PAPER4}. We estimate a total radiation load of $\sim$20~W to be sunk towards the first stage of the PTs, when looking at the 300~K laboratory background. This will decrease to $\sim$14~W on site, when the window will receive radiation from the low-emissivity sky and the absorbing forebaffle, resulting in an improvement in the performance of the cryogenic system. 

\subsubsection{Loads on the 3-K stage}

The 3-K stage serves as a thermal anchor for the electrical harnesses that are connected to the feedhorn array switches, the HWP rotator, and the detectors.
The 3-K stage also aims to limit the transmission, towards the 1-K stage, of the radiation re-emitted in the optical chain by the 40-K filter, and of the radiation emitted by the 40-K shield. The first contribution is limited by means of low-pass metal-mesh filters, thermally connected to the 3-K stage; these reflect back most of the out-of-band radiation, absorbing only a small fraction (a few percent). Their re-emission is reduced, being thermally sunk at 3~K. The other important function of the 3-K stage is to sink the heat produced by the evaporation refrigerators when heating the cryopumps and condensing $^3$He and $^4$He. This heat load can be significant ($\sim$0.5W). If added to the other heat loads, it can raise the temperature of the 3-K stage enough to render cryogens condensation inefficient. For this reason, we have decided to dedicate one PT mainly to cool the 3-K shield, and the other one to cool the flanges of the evaporation refrigerators, where the condensation of $^4$He and $^3$He happens (see figure \ref{fig:PTs}). 

As shown in table \ref{tab:loadsmain}, we have designed the cryostat to reduce the total heat load on the 3-K stages well below the  cooling power of 0.9~W which would be available at 4.2~K for each PT. This is particularly true for PT2, which is devoted to the EVRs. In this way the second stage of PT2 can achieve the low ($\sim$3.1~K) temperature necessary for efficient condensation of $^4$He in the two evaporation refrigerators. 

\begin{table}[htbp]
\centering
\caption{\label{tab:loadsmain} Calculated heat loads in the main QUBIC cryostat in the final instrument configuration. }
\smallskip
\begin{tabular}{|l|l|r|}
\hline
stage & origin & power \\
\hline
40-K & radiation from outer shell (through MLI) & 9.2~W \\
40-K & radiation from window (site / laboratory) & 5~W/11.3~W \\
40-K & conduction from G10 struts & 1.0~W \\
40-K & conduction from harness & 0.7~W \\
40-K & dissipation from amplifiers & 0.4~W \\
\hline
40-K & total load on PT1 \& PT2 (site / laboratory) & {\bf 16~W  / 23~W} \\
\hline
3-K PT1 & radiation from 40-K shield & 80~mW \\
3-K PT1 & radiation from optical system & 3~mW \\
3-K PT1 & conduction from G10 struts & 60~mW \\
3-K PT1 & friction from HWP rotator (average) & 10~mW \\
3-K PT2 & conduction from harness & 11~mW \\
\hline
3-K PT1 & total load on PT1 & {\bf 164~mW} \\
\hline
3-K PT2 & load from EVRs (when cycling) & 500~mW \\
\hline
3-K PT2 & total load on PT2 (when cycling the EVRs) & {\bf 500~mW} \\
\hline
\end{tabular}
\end{table}

Due to the significant heat flow when the EVRs are cycled, the thermal resistance between the cold head of PT2 and the EVRF must be minimized. In fact, there is no heat switch separating the two components, and we used copper braids for heat transfer as detailed in \S \ref{par:40K3Kstages}. In this way, even with 0.5~W dissipated in the cryopumps, the temperature rise of the EVRF with respect to the cold end of PT2 is expected to be $<0.2$~K.

\section{1-K and 0.3-K stages}\label{par:1K03K}

\begin{figure}
\centering
\includegraphics[width=1\linewidth]{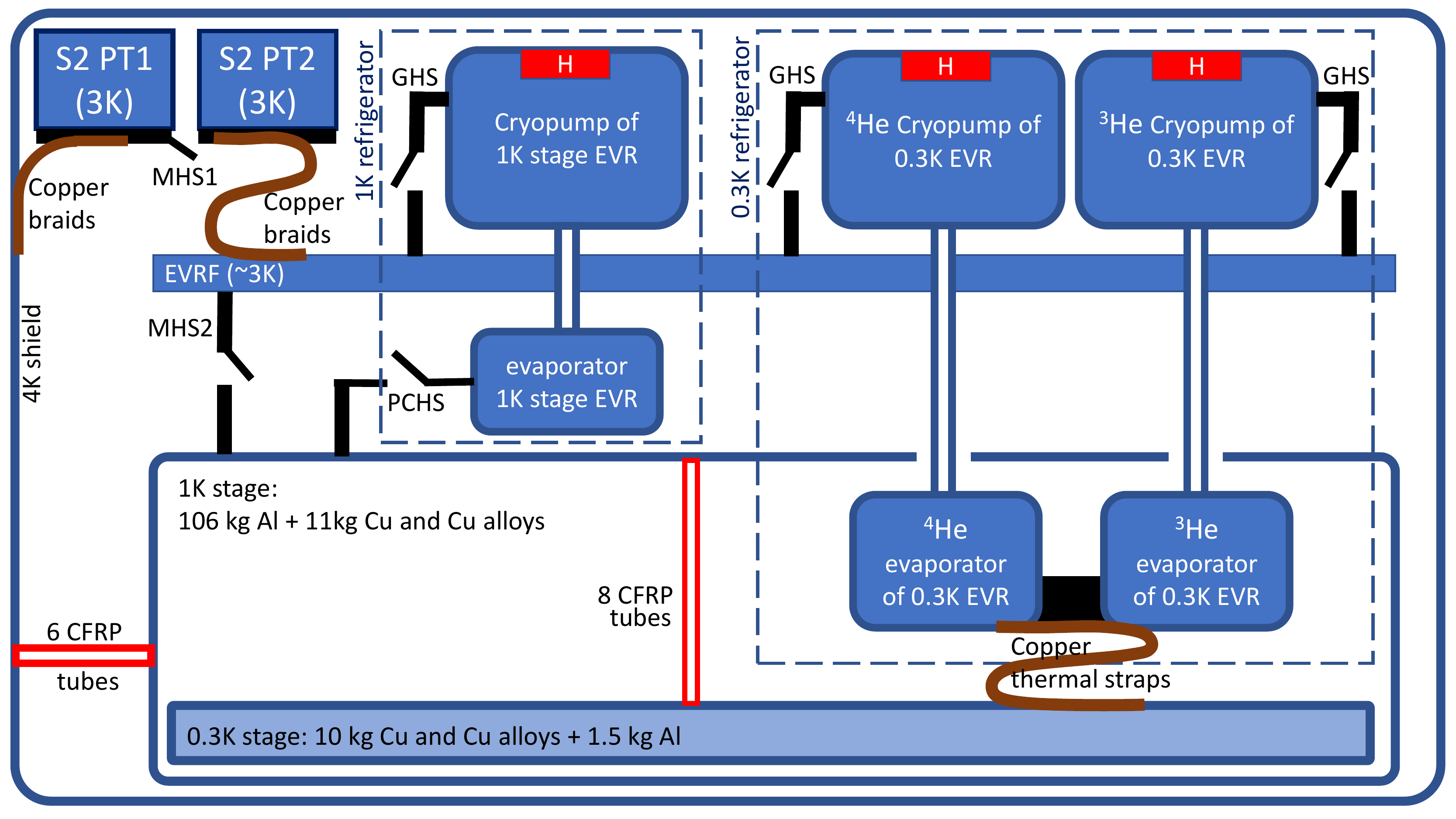}
\caption{Thermal architecture of the 1-K and 0.3-K refrigeration stages. The evaporation refrigerators flange (EVRF) is cooled at $\sim$3~K by the second stage of pulse tube 2 (S2PT2), and supports the 1-K stage evaporation refrigerator and the 0.3-K stage evaporation refrigerator. For pre-cooling purposes, the second stage of PT1 can be connected via a mechanical heat switch (MHS1) to the the second stage of PT2, to have the two PTs operating in parallel to pre-cool the EVRF, and also to keep it cold when the cryopumps are heated. In addition, the 1-K stage can be thermally connected to the EVRF via a second mechanical heat switch (MHS2). The 0.3-K stage, instead, is pre-cooled through convection of the $^3$He and $^4$He gases present in the 0.3-K refrigerator. The evaporator of the 1-K refrigerator cools the 1-K stage via a passive convection heat switch (PCHS, see text for a description of its operation). The 0.3-K stage is cooled first to $\sim$1~K by the $^4$He evaporator of the 0.3-K refrigerator (also allowing for efficient condensation of the $^3$He in the $^3$He evaporator, since the two evaporators are in thermal contact), and then to 0.3~K by the evaporator of the 0.3-K refrigerator. All the cryopumps can be heated, to desorb the gas, by suitable electrical heaters (H). They can also be cooled to adsorb the gas and reduce the vapor pressure above the liquid in the evaporator by means of gas-gap heat switches (GHS).
\label{fig:subKarchitecture}} 
\end{figure}

The heart of the QUBIC instrument (detectors, filters and preamplifiers) must be cooled to 0.3~K, and the surrounding vital parts must be cooled to 1~K, to reduce the heat load on the detectors and on the 0.3-K refrigerator. This is achieved by means of two self-contained EVRs \cite{2018SPIE10708E..3VM}. 
Their general configuration 
is summarized in figure \ref{fig:subKarchitecture}. 

\subsection{1-K stage}

\subsubsection{Thermo-mechanical design}\label{ss:tmd}
The 1-K section of the instrument is massive, and poses a significant challenge both for the precooling at 3~K, by means of the PT, and for the further cooling at 1~K, by means of the dedicated 1-K evaporation refrigerator. In detail, this section of the instrument is composed of 130~kg of 6061 aluminum alloy, 11~kg of AISI104 stainless steel, 10~kg of OHFC copper, 4~kg of brass, and 10~kg of Stycast 2850 impregnated with Eccosorb HR10 (covering the 1-K stage walls), for a total of 165~kg. 

\begin{figure}
\centering
\includegraphics[width=1\linewidth]{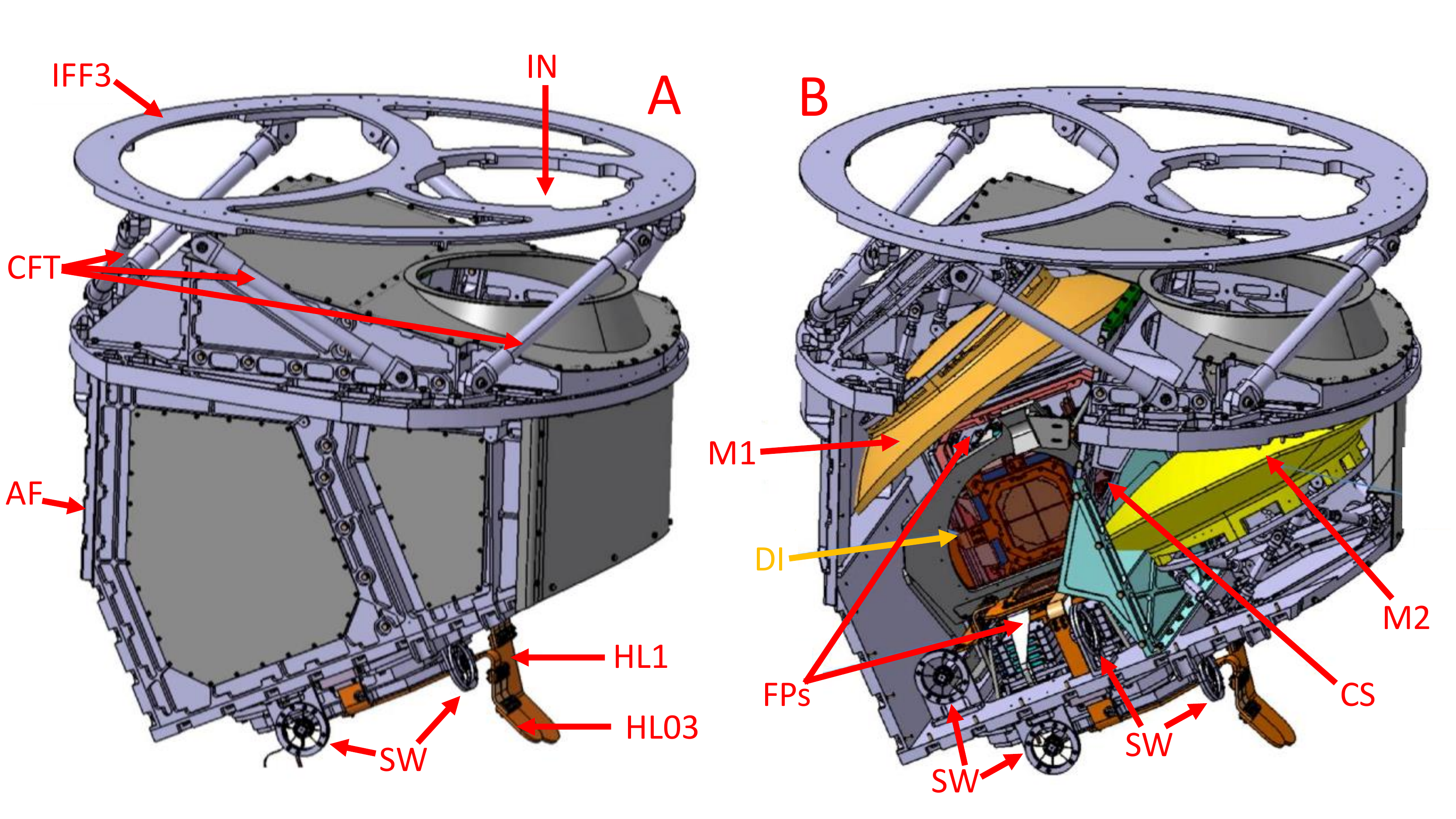}
\caption{The 1-K section of the QUBIC instrument. On the {\bf left} (A), the aluminum frame (AF) supporting the subsystems is shown, with all the shields in place. The optical beams enter the 1-K section through the top aperture (IN). The 1-K section is suspended under the 3-K stage (interface flange IFF3) by means of six carbon-fiber-reinforced-plastic tubes (CFT) providing the necessary thermal insulation. The gold-plated OFHC copper heat links to the 1-K evaporation refrigerator (HL1) and the 0.3-K evaporation refrigerator (HL03) are visible in the bottom part of the section. These are supported by stiff, low conductivity support ``wheels'' (SW)}. On the {\bf right} (B) part of the frame and of the radiation shield panels have been removed, to show the internal components: the off-axis Al mirrors of the beam combiner (M1 and M2), the cold stop (CS), the dichroic filter and beamsplitter (DI) and the two focal planes, with the detector arrays (FPs). \label{fig:oneK} 
\end{figure}

The 1-K stage is supported from the 3K stage by means of six carbon-fiber tubes forming a stiff hexapod, as shown in figure \ref{fig:oneK}. The tubes are $\sim$0.33-m long, 1-mm thick, with 30~mm OD. Their composition is 60$\%$ carbon fiber (by volume) and 40$\%$ resin (DP406), for a conductivity of 12.7~mW/m between 5~K and 1~K. This results in  a heat load on the 1-K refrigerator of 0.32~mW. The carbon-fiber tubes are glued to aluminum rod ends with epoxy glue (Ablestic 286 white).  The calculated heat loads on the 1-K stage are summarized in table \ref{tab:1Kloads}.
\begin{table}[htbp]
\centering
\caption{\label{tab:1Kloads} Calculated heat loads on the 1-K and 0-3K stages of the QUBIC cryostat.}
\smallskip
\begin{tabular}{|l|l|r|}
\hline
stage & origin & power \\
\hline
1-K & CFRP tubes & 0.32~mW \\
1-K & Instrumentation wire harness & 15~$\mu$W \\
1-K & radiation from optical chain & 1~$\mu$W \\
1-K & radiation from 3-K shield & 6~$\mu$W \\
1-K & off-conductivity of convective heat switches & 0.33~mW \\
\hline
1-K & total load  &  {\bf 0.65~mW} \\
\hline
0.3-K & CFRP and stainless steel tubes & 17~$\mu$W \\
0.3-K & Instrumentation wire harness & 6~$\mu$W \\
0.3-K & radiation from optical chain & 1~$\mu$W \\
0.3-K & radiation from 1-K shield & $\ll 1~\mu$W \\
\hline
0.3-K & total load  & {\bf 25~$\mu$W} \\\hline
\end{tabular}
\end{table}

In order to achieve a reasonable time to pre-cool down to 3-K, we avoid the use of exchange gas (which is dangerous for some optical filters and would degrade the performance of the multilayer radiation insulation), and use mechanical and convective heat switches. The mechanical heat switch MHS2 (see figure \ref{fig:PTs}) is closed at the beginning of the cool-down procedure, to bypass the thermal insulation between the 1-K stage and the 3-K stage. The heat switch is based on a commercially available electro-mechanical device \cite{entropy}, a clamp based on two opposite worm screw jacks operated by a cryogenic step motor.  

\subsubsection{1-K evaporation refrigerator}\label{par:1K}

The specific heat of the 1-K section of the instrument is such that 18~MJ of heat must be removed to cool it down from 300~K to 1~K. While most of the heat is removed by the PT, the final cooldown from 4~K to 1~K is obtained by means of the 1K evaporation refrigerator, and implies the removal of 60~J of heat. The 1-K EVR has been designed \cite{2016SPIE.9912E..66M} to cope with this heavy requirement, which implies a large charge of $^4$He gas (1.5~mol) in the system, stored at high pressure at room temperature (75~bar at 300~K). A heat switch has been inserted between the evaporator and the 1-K stage to limit the required $^4$He charge , and operating a sequence of recycles of the refrigerator until  
the full charge of liquid $^4$He can be used to keep the 1-K stage at a stable temperature of $\sim$ 1~K for 24~hours, under the heat loads specified in table \ref{tab:1Kloads}. 
An efficient film-breaker has been inserted in the tube connecting the evaporator pot to the condenser to avoid the high conductivity of the superfluid $^4$He film rising along the tube wall (see e.g. \cite{1989AA...226..357M, 2019Cryo..102...45M}). 

A special passive heat switch has been developed to avoid the heat load on the 1-K stage when the cryopump is heated at $>$30~K to desorb the convection gas. The switch (PCHS in figure \ref{fig:subKarchitecture}) consists in a simple convective circuit, similar to the one described in \S \ref{ss:tmd}, without the cryopump. 
Convection 
is naturally activated when the top of the switch is colder than the bottom. The top of the switch is connected to the evaporator pot of the 1-K EVR, while the bottom is connected to the 1-K stage. In this way, when the 1-K EVR is at the low ($\sim$1~K) operating temperature, and the 1-K stage is warmer, convection is active, the switch is closed, and the EVR cools the 1-K stage. When the 1-K EVR is being cycled, with warm gas desorbed by the cryopump and warming up the evaporator pot, or with unpumped liquid $^4$He in the evaporator pot, convection is stopped, the switch is open, and the warm temperature of the evaporation pot does not warm-up the 1-K stage. This switch has been charged at a pressure of 10~bar at room temperature, and achieves a conductivity of 10~mW/K in the ON-state at 1~K. With $\sim$0.7~mW of heat flow through the switch, we expect a thermal gradient of $\sim$70~mK between the 1-K stage (at $\sim$0.97~K) and the 1-K evaporator pot (at $\sim$0.90~K). In the OFF-state, a small residual conductivity is present due to the superfluid film. This issue has no practical impact on the operation of QUBIC, and could be solved using $^3$He as the convective gas in place of $^4$He. 

\begin{figure}
\centering
\includegraphics[width=0.9\linewidth]{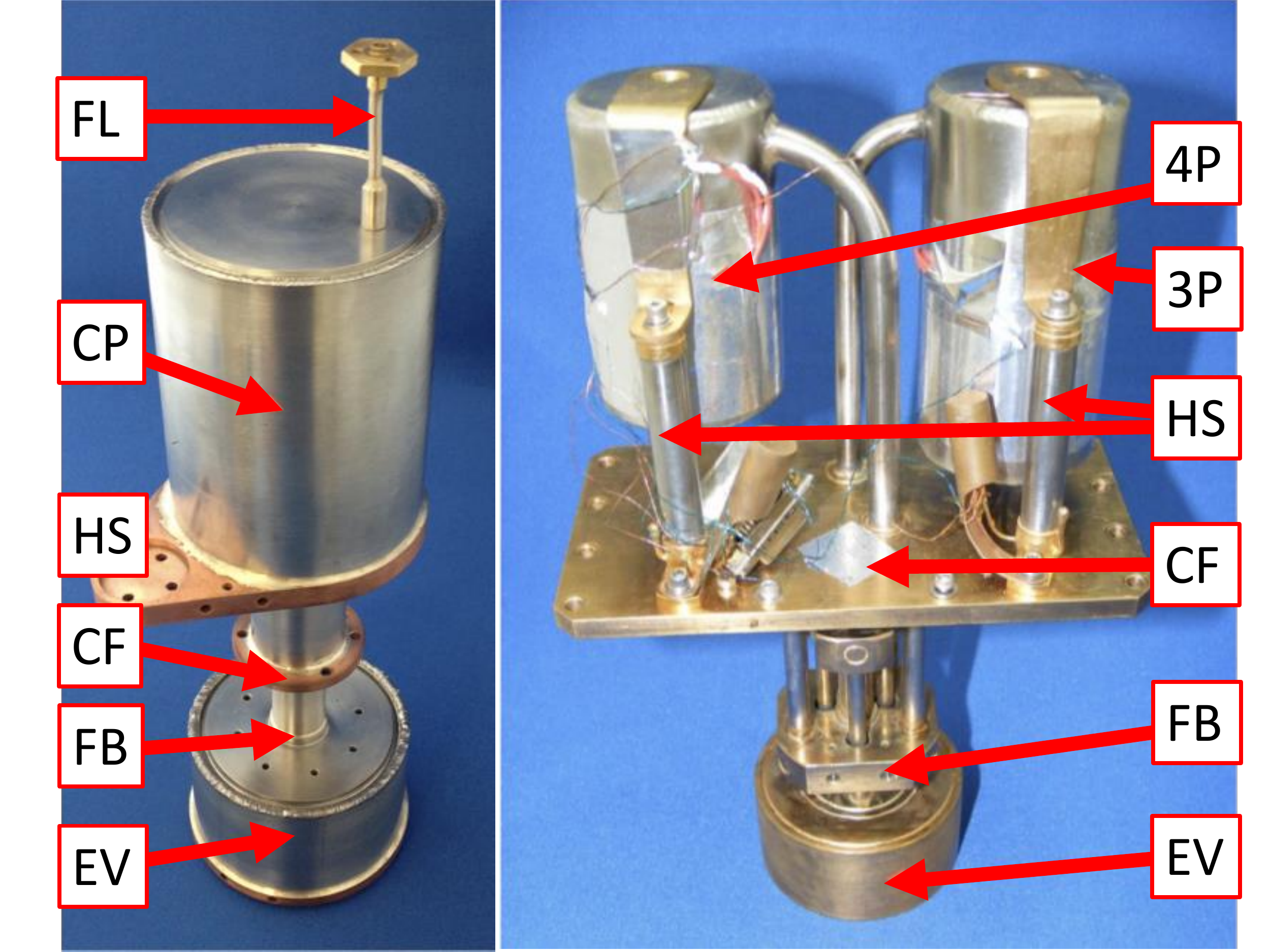}
\caption{{\bf Left}: The 1-K evaporation refrigerator of QUBIC. From top to bottom: the fill line (FL), the cryopump (CP), the heat switch flange (HS), the condensation flange (CF), the film breaker (FB, not visible) and the evaporator pot (EV). The refrigerator is $\sim$350~mm tall. {\bf Right}: The 0.3-K evaporation refrigerator of QUBIC. From top to bottom: the $^4$He cryopump (4P), the $^3$He cryopump (3P), the gas-gap heat switches (HS), the condensation flange (CF), the film-breaker assembly (FB) and the evaporator pots (EV). The refrigerator is $\sim$250~mm tall.
\label{fig:1K03K}} 
\end{figure}

A picture of the 1-K EVR is shown in figure \ref{fig:1K03K}. The main challenge for this refrigerator is the relatively high temperature provided by the PT for the condenser flange (EVRF in figure \ref{fig:PTs}). In fact the expected condensation efficiency (fraction of the total $^4$He gas charge which is condensated in liquid form in the evaporator pot) is a strong function of the temperature. During the qualification test, under the nominal heat load (table \ref{tab:1Kloads}), the measured hold time was $\sim$30~h for condensation at 2.7~K, decreasing to $\sim$15~h for condensation at 3.7~K. In the QUBIC cryostat the condensation temperature provided by S2PT2 is $\sim$3.1~K.

\subsection{0.3-K stage}

\subsubsection{Thermo-mechanical design}
\begin{figure}
\centering
\includegraphics[width=0.48\linewidth]{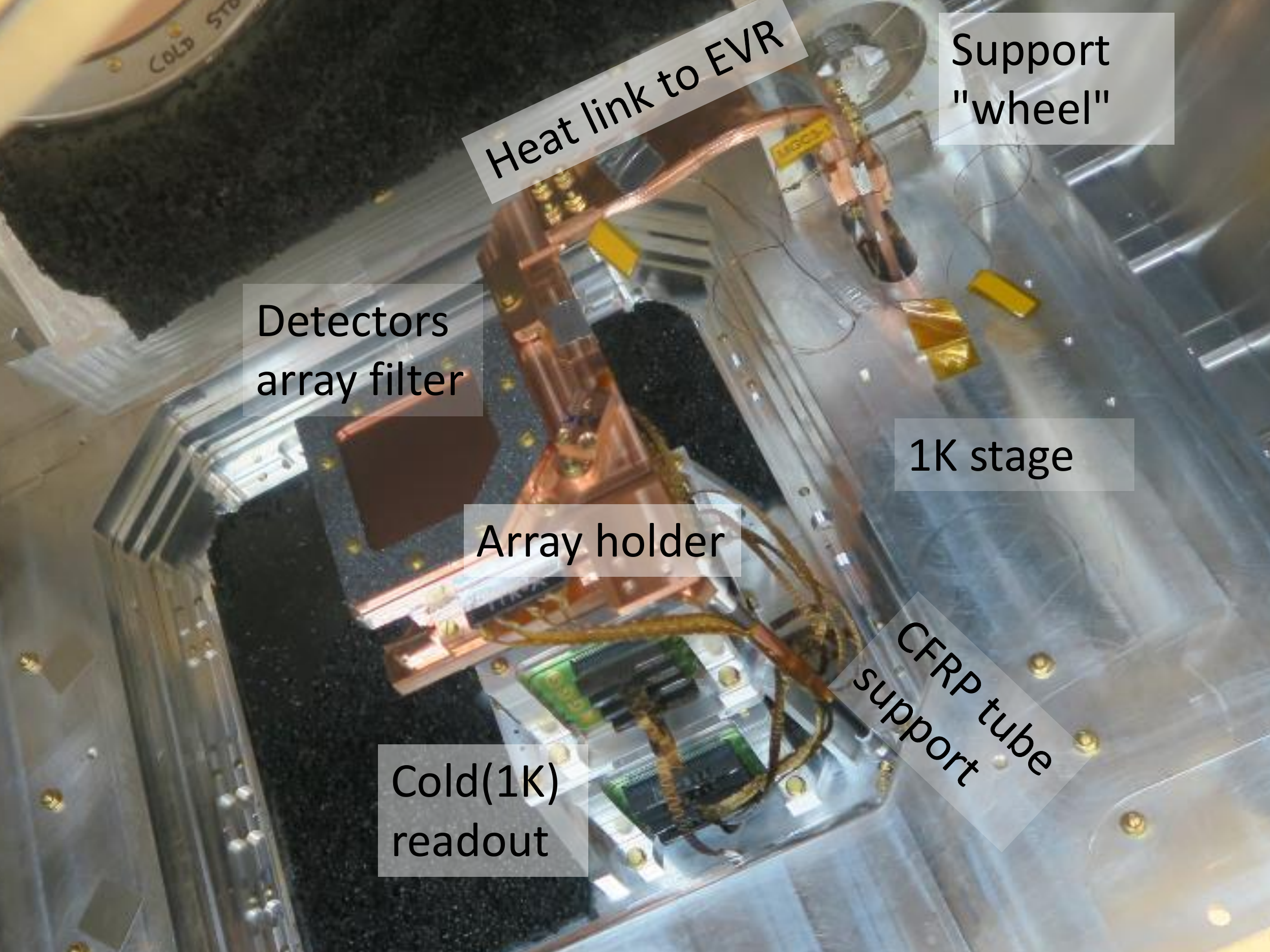}
\includegraphics[width=0.48\linewidth]{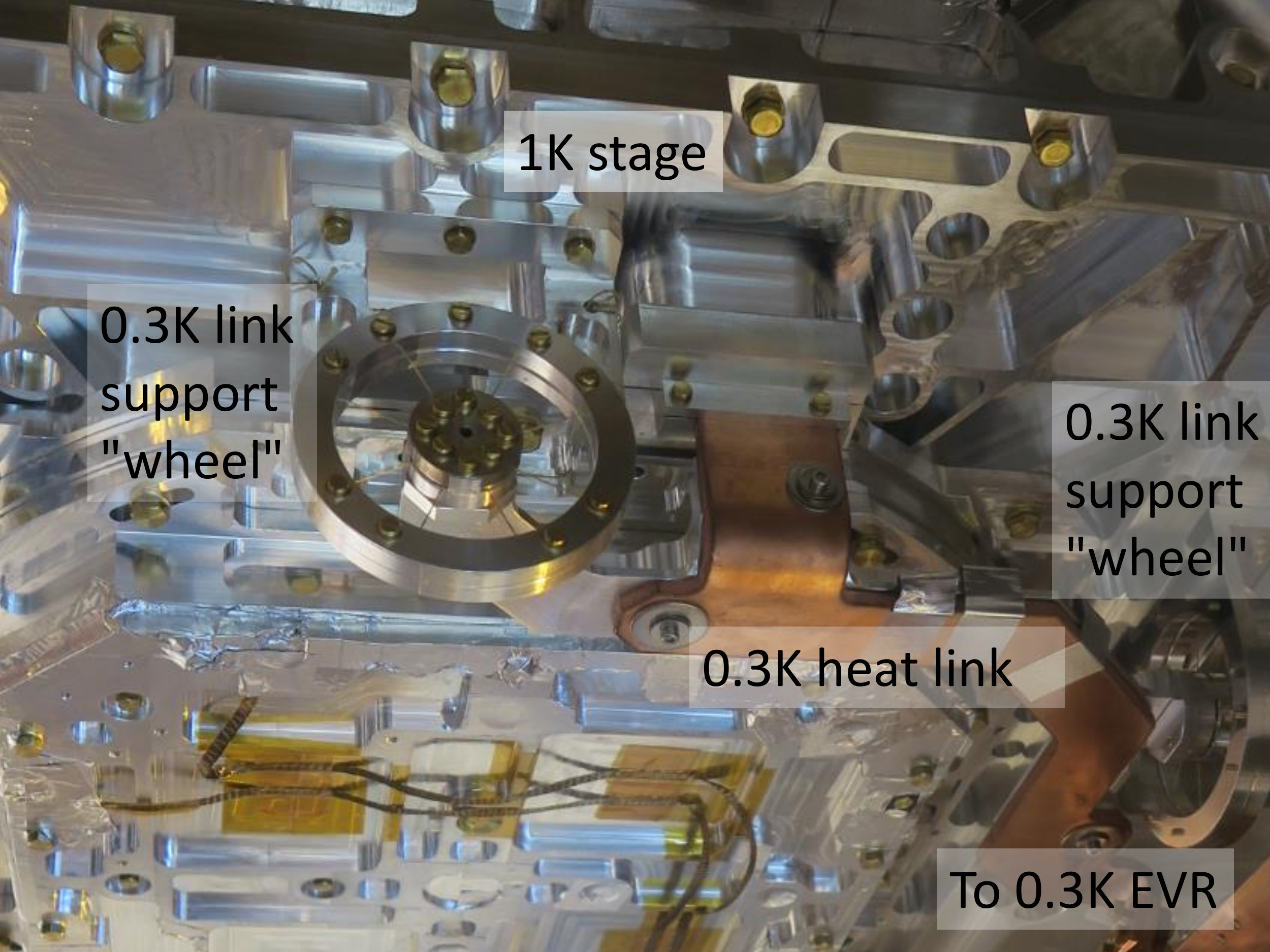}
\caption{{\bf Left}: Detector array module mounted in the QUBIC cryostat. This module is one of the 4 subarrays filling the transmitted focal plane. It is cooled by the 0.3-K EVR through a OFHC Cu heat link and supported by 3 CFRP tubes.  {\bf Right}: Continuation of the OFHC Cu heat link towards the evaporation pot of the 0.3-K EVR, and its support system based on 1~mm diameter stainless steel tubes organized in ``wheels'' (see also figure \ref{fig:oneK}).  \label{fig:FP}} 
\end{figure}

The two focal plane arrays are filled with TES bolometric detectors. The bias power and the radiative background set the temperature of the absorber and the superconducting thermistor in the middle of the superconductive transition ($\sim$500~mK), starting from a base temperature of $<$350~mK. This sets the requirement for the steady temperature to be reached by the 0.3-K EVR. The other requirements come from the suspended mass to be cooled, and the parasitic heat load from the supports and the residual radiation. The total mass of the two focal plane assemblies (see figure \ref{fig:FP}), including the optical filters, the wafers, the wafer holders and the connectors, is $\sim$1~kg. They are supported by a set of CFRP tubes (3 for the TD, 8 in the FI), that are $\sim$110-mm long, and have 5.5~mm OD, 3.5~mm ID. Each focal plane assembly is connected to the evaporation pot of the 0.3-K EVR through an  OFHC copper link (see figure \ref{fig:FP}). This adds $\sim$0.5~kg to the total mass to be cooled. The link is suspended over the 1-K frame by means of 3 support ``wheels'' each with 3 spokes made in stainless steel tube of 1~mm OD, 0.9~mm ID and 19.5-mm long, as seen in figure \ref{fig:FP}.  The expected heat loads are listed in table \ref{tab:1Kloads}.

\subsubsection{0.3-K evaporation refrigerator}
A commercially available double-stage $^3$He/$^4$He evaporation refrigerator \cite{chase} fits the requirements described above. This features two separate cryopump-condenser-evaporator circuits, with the two evaporators in close thermal contact (see figure \ref{fig:subKarchitecture}). One of the two circuits is filled with 0.71~mol of $^4$He and operates as the 1-K EVR described in \S \ref{par:1K}. The other one is filled with 0.36~mol of $^3$He. 

The purpose of the first EVR is to cool the evaporator pot to 1-K for the time required for complete condensation of $^3$He in its evaporator pot. After condensation of $^3$He is completed, and the $^4$He is exhausted, the cryopumping of the $^3$He circuit is activated, and the evaporator pot reaches a temperature $\sim$0.3~K. This is kept constant until the L$^3$He is completed exhausted, allowing the operation of the detectors in a very stable temperature environment. After complete exhaustion of the L$^3$He, the cycle is repeated. For this EVR, the operation of the two cryopumps is controlled by two gas-gap heat switches \cite{bookPiccirillo}. A picture of the system is reported in figure \ref{fig:1K03K}. With the use of a dedicated 1-K refrigerator, complete condensation of the full $^3$He charge is possible, overcoming the problem of the relatively high ($\sim$3.6~K) temperature provided by the pulse-tube for the condenser flanges. In the qualification test, the 0.3-K EVR reached a temperature of 336~mK under the nominal load (table \ref{tab:1Kloads}), with a hold time of 90~h, and negligible instability ($<$1~mK/$\sqrt{Hz}$ at 0.01~Hz).

\section{Test and Validation}

\subsection{Performance of the 40-K and 3-K stages}
\subsubsection{Pre-cooling time}
Given the large mass and heat capacity of the instrument parts to be cooled, the finite cooling power of the cryocoolers, and the extension of the thermal links from the PT cold-heads to the different instrument parts, the time required to thermalize the 40-K and 3-K stages is not negligible. 

\begin{figure}
\centering
\includegraphics[width=1\linewidth]{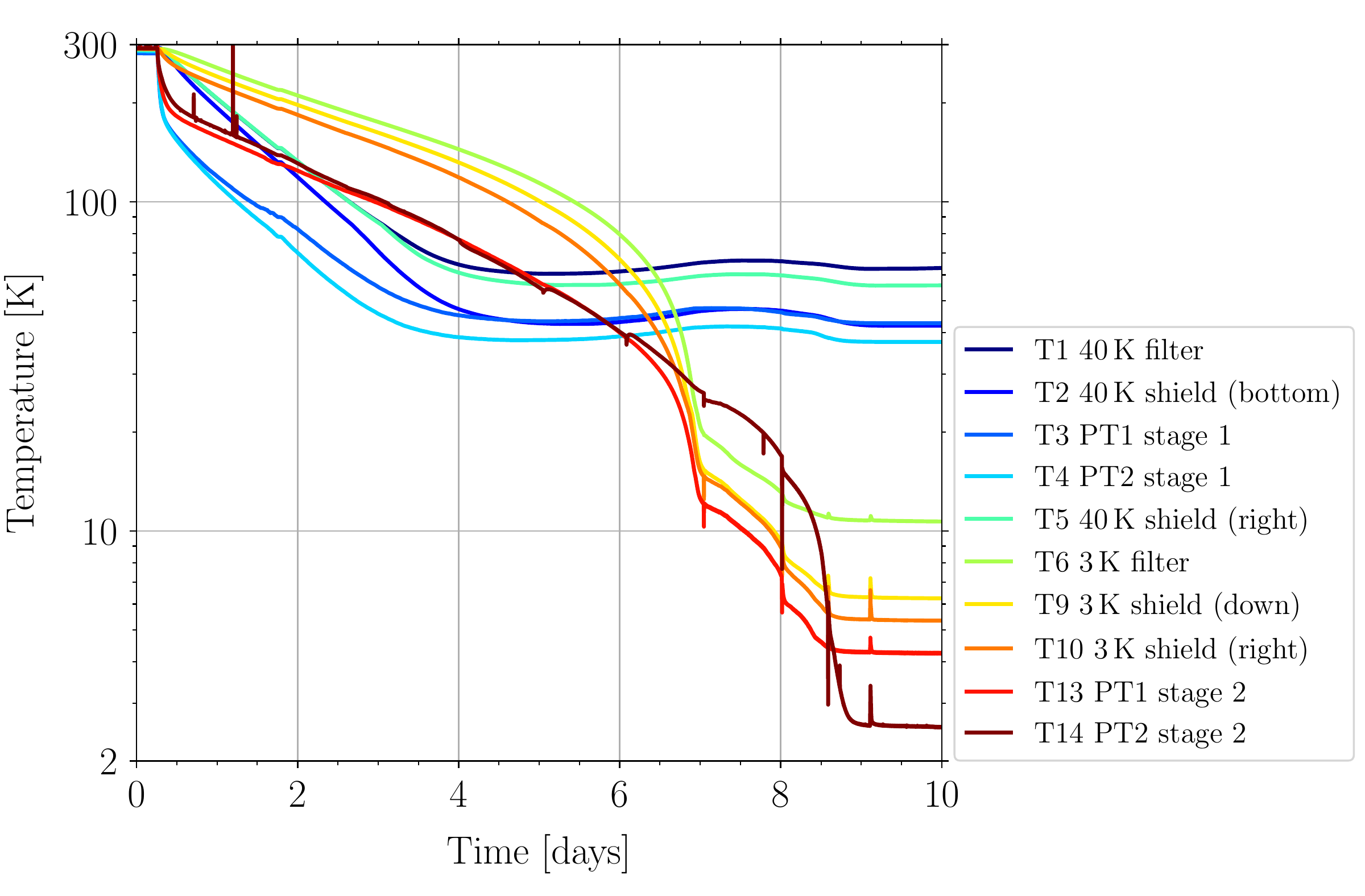}
\caption{Temperatures measured by different thermometers in the 40-K and 3-K stages, during the cooldown of Oct. 25$^{th}$, 2019. \label{fig:maincooldown}}
\end{figure}

The main cryostat (outer shell, cryocoolers, 40-K and 3-K stages) was first tested empty, in Rome, immediately after assembly, with all the windows blanked with metal flanges. The asymptotic temperatures of (40$\pm$1)~K and (3.1$\pm$0.5)~K, for the first and second stages respectively, were achieved in $\sim$2~days. After this first commissioning test, the system was delivered to Paris where the 1-K and 0.3-K stages and the instrument insert were integrated. The complete system has been cycled a number of times since, for different configurations and many months of continuous operation. 
In figure \ref{fig:maincooldown} we report the temperatures sensed by the 40-K and 3-K stages thermometers, during the cooldown started on Oct. 25$^{th}$, 2019. 

The cooling time to reach the asymptotic temperatures of the 40-K and 3-K cold heads is $\sim$8.2~days. This is consistent with the prediction of a simple thermal model including the heat capacity of $\sim$165~kg of metals to be cooled by the PTs and taking into account the heat loads and the cooling power of the PTs, and the variation of all these quantities with temperature. The asymptotic temperatures of the PT heads are (42$\pm$1)~K, (38$\pm$1)~K, (4.2$\pm$0.1)~K, (2.5$\pm$0.1)~K, for the PT1 and PT2 40-K stages and the PT1 and PT2 3-K stages, respectively. These temperatures were expected to be higher than in the commissioning test, due to the heavy heat load through the window and the filters, but are anyway satisfactory. 

Comparing the measured asymptotic temperatures to the nominal capacity chart of the PTs, one concludes that the heat load on each of the two 40-K stages is $\sim$30~W. A total of 60~W is significantly more than expected on the 40-K stages (see table \ref{tab:loadsmain}). The extra load might be due to the presence of radiation leaks in the MLI blanket around the support straps, and radiation leaks in the optical filter train following the optical window. The MLI blanket will be refurbished and improved after shipment to Argentina, and the filters will be fixed.

For the 3-K stages, the 2.5~K asymptotic temperature of PT2 implies a $\ll$1~W heat load, as expected when the cryopumps are not operated. The 4.2-K asymptotic temperature of PT1 implies a $\sim$1~W heat load. This is also more than expected from table \ref{tab:loadsmain}, possibly due to radiation leaks in the optical filters train as suggested by visual inspection of some of the filters, which will be fixed.

Significant temperature gradients are present between the PT heads and the other parts of the stage, for both stages, as evident from the temperature measurements plotted in figure \ref{fig:maincooldown}. For the 40-K stage, the 40-K filter area runs 20~K hotter than the PT head, due to the significant heat load from the window, while the shield area opposite to the PT head runs 15~K hotter than the PT head. These gradients are due to the finite conductivity of the Al shield and the important radiative heat load from the 300-K shell, but are still compatible with the operation of the system. For the 3-K stage, the filter area is $\sim$6~K hotter than the PT head, while the different areas of the 3-K shield (underneath and opposite to the PT) are reasonably thermalized.

\subsubsection{Response to non-static heat loads}
The 3-K stage is subject to non-static heat loads when the HWP mechanism is operated and when the cryopump heaters in the EVR are activated. 

\begin{figure}
\centering
\includegraphics[width=0.65\linewidth]{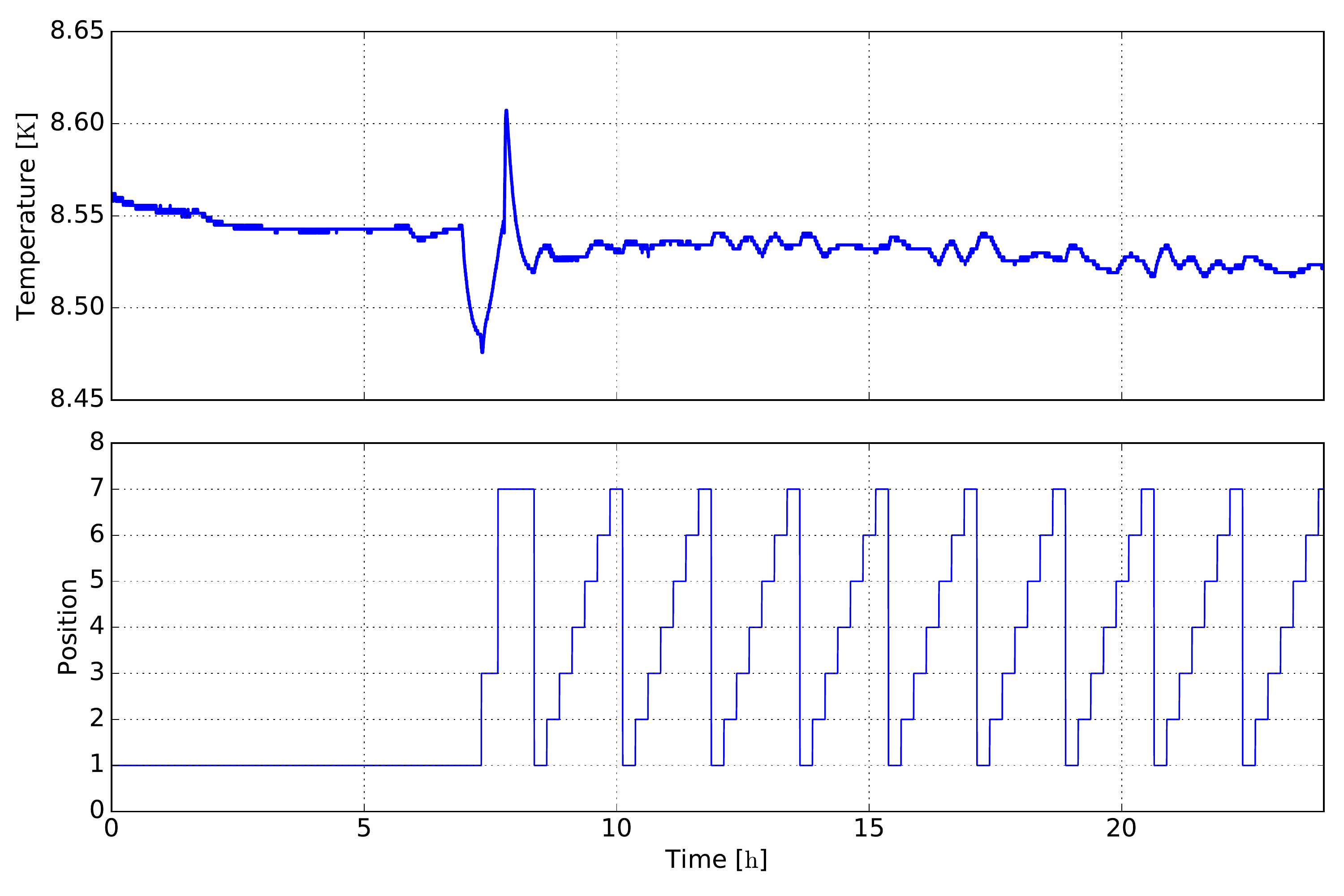}
\caption{{\bf Top:} temperature measured by the thermometer closest to the HWP rotation mechanism (3-K stage, near filter). {\bf Bottom:} HWP position (each of the positions 1-7 is separated from the next one by \SI{15}{\degree} of HWP rotation). The position is recorded only at the end of the movement. \label{fig:friHWP}}
\end{figure}

The effect of the rotating HWP is modest (in fact the system has been optimized with friction minimization in mind \cite{2020AA.QUBIC.PAPER6}). 
In figure \ref{fig:friHWP} we show a record of the filter ring temperature during a time period when the HWP was activated. The filter ring thermometer is the thermometer that is closest to the HWP. The typical temperature variation at the 3-K filter ring, associated with large motions of the nearby mechanism, is 30~mK. During regular scans (15$^\circ$ steps every 15 minutes), the motion of the mechanism produces typical temperature fluctuations of $\sim$5~mK {\it rms}, correlated with the motion. All the other thermometers are insensitive to the motion of the HWP mechanism.
\begin{figure}
\centering
\includegraphics[width=1\linewidth]{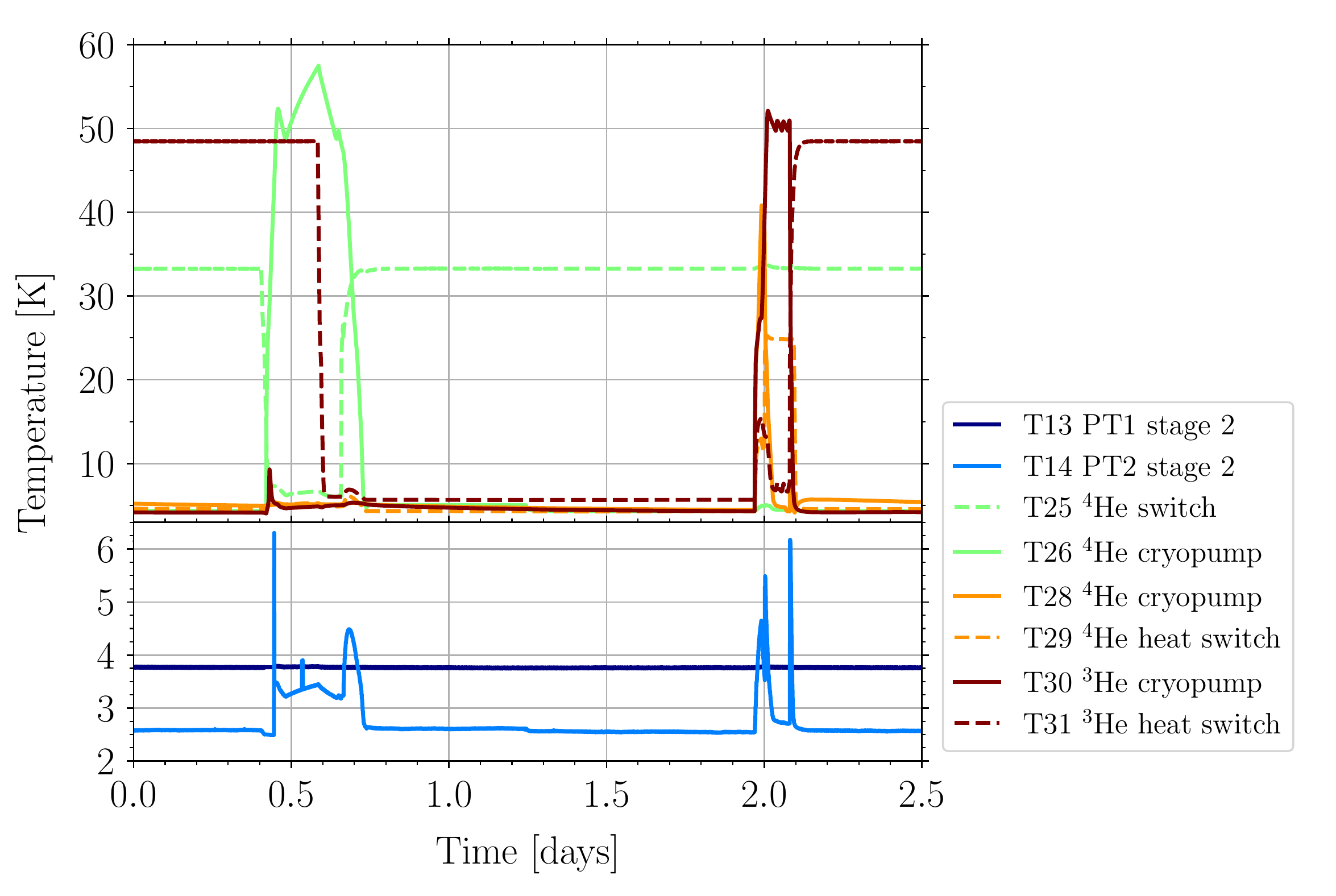}
\caption{Temperature of the different thermometers in the 40-K and 3-K stages during sample heating cycles of the EVRs cryopumps, desorbing warm $^3$He and $^4$He.  \label{fig:cryopumploads}}
\end{figure}

The activation of the cryopump heaters in the EVRs, resulting in the desorption of warm gas, produces a large heat load on the 3-K stage, with a consequent warming of the EVRF and the cold head of PT2, as visible in figure \ref{fig:cryopumploads}. Depending on the cryopump heating profile, the temperature rises by $\sim$1~K. Considerable effort has been spent to maximize the heat conductivity between the condensation point of the 0.3-K and 1-K EVRs and the cold head of PT2, as described in \S \ref{par:40K3Kstages}. The desorbed gases in the EVRs condense at around 3.6~K, a temperature providing sufficient condensation efficiency (see \S \ref{par:1K}, \S \ref{par:1Kperformance}, and \S \ref{par:0p3Kperformance}).

\subsection{Performance of the 1-K stage}\label{par:1Kperformance}

\begin{figure}
\centering
\includegraphics[width=1\linewidth]{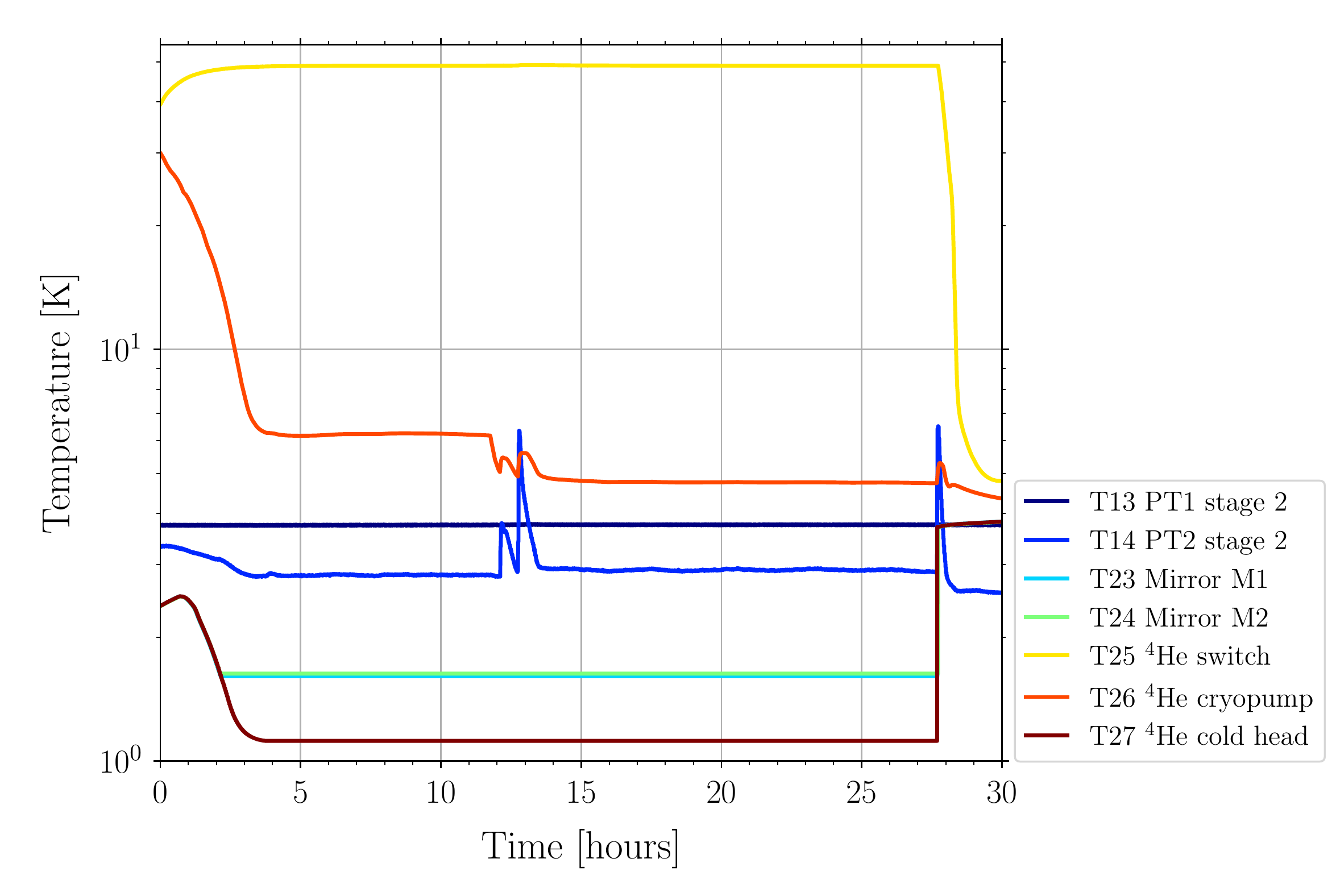}
\caption{Temperature of the cold head of the $^4$He EVR (lower trace) and of related components during one EVR cycle. The instabilities at $\sim$12-13~hours are mainly due to the QUBIC instrument operations carried out at that time. The hold-time is $\sim$24~hours. \label{fig:1Kperformance}}
\end{figure}

During preliminary commissioning in a dedicated cryostat, the capacity of the 1-K refrigerator was measured to be 25~J, with condensation at 3.6~K, reaching and maintaining a cold head temperature of 0.9~K with the 1~mW load expected during operation in the QUBIC cryostat for more than 15 hours. After commissioning, the refrigerator was installed in the QUBIC cryostat as described in \S \ref{par:1K03K}. The condensation temperature is about 3.1~K, and the measured capacity of the 1-K refrigerator inside QUBIC is about 51~J. 
We have verified that this refrigerator allows us to cool the heavy 1-K stage to $\sim$1.1~K, and maintain it at this temperature for more than 24 hours, as evident from figure \ref{fig:1Kperformance}. 

\subsection{Performance of the 0.3-K stage}\label{par:0p3Kperformance}

\begin{figure}
\centering
\includegraphics[width=14cm, height=9cm]{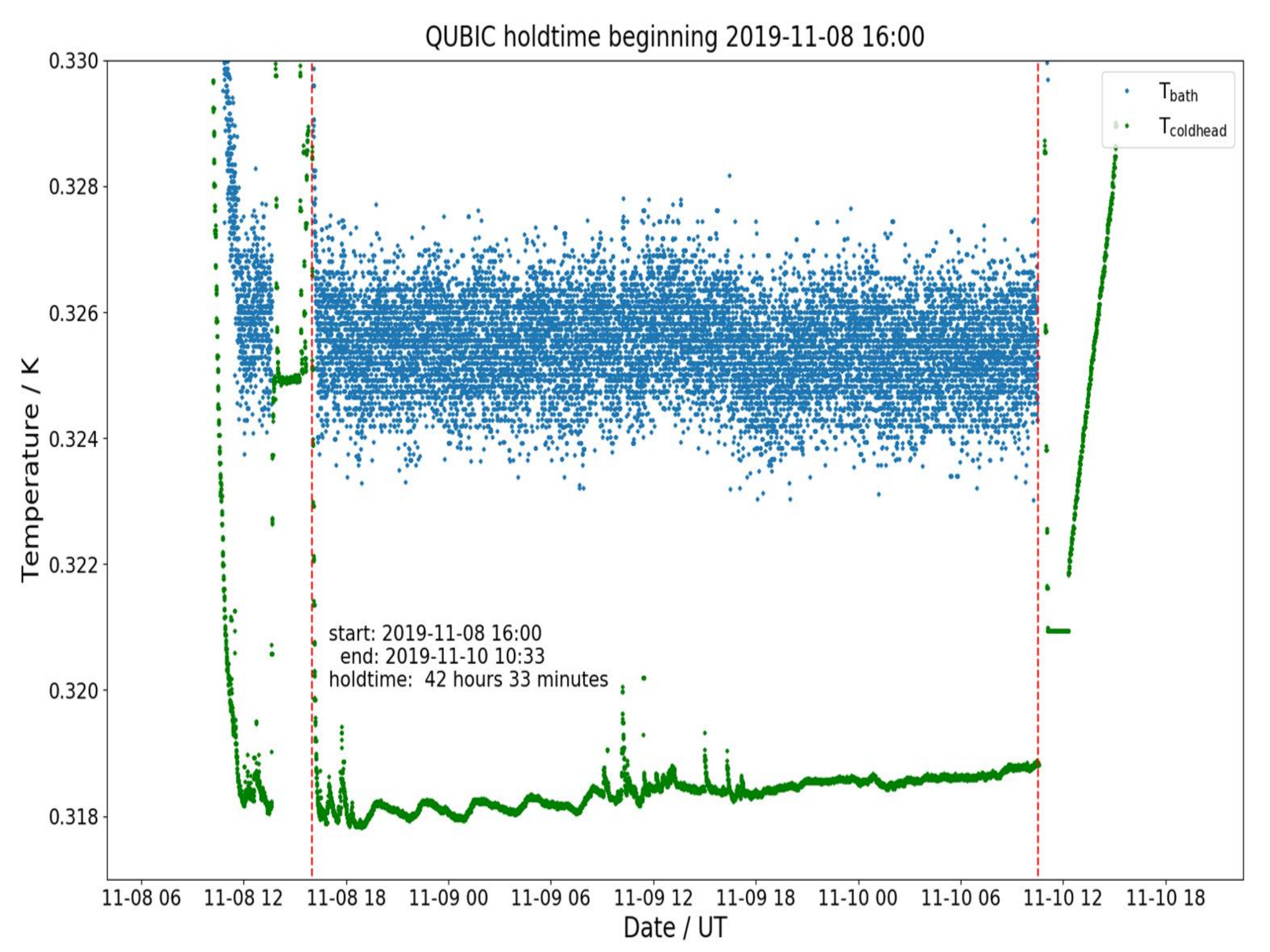}
\caption{Temperature of the cold head of the $^3$He EVR (lower trace) and of the detector flange (upper trace) during the operation of the $^3$He EVR. The small instabilities are mainly due to the QUBIC instrument operations carried out at that time. The hold-time exceeds 42~hours.  \label{fig:03Kperformance}}
\end{figure}

During preliminary commissioning in a test cryostat, the capacity of the 300-mK refrigerator was measured to be 8.1~J (hold time of 90~hours with 25~$\mu$W load giving a cold head temperature of 339~mK). After commissioning, the refrigerator was installed in the QUBIC cryostat as described in \S \ref{par:1K03K}.
During these laboratory tests, a 9\% transmission neutral density filter was inserted in the optical path, at a temperature of 1~K, to avoid saturation of the TES detectors (see \cite{2020AA.QUBIC.PAPER3} for details). From the cryogenic point of view, the only effect of this filter is to reduce the load on the 0.3-K stage from the optical chain, which is already a minor contribution (see table \ref{tab:1Kloads}).
A plot of the temperature of the cold head and of the detectors flange is reported in figure \ref{fig:03Kperformance}.
The performance of the $^3$He refrigerator in the QUBIC cryostat (in the TD configuration) is consistent with that measured in the preliminary commissioning. The operating temperature of 318~mK at the cold head has been achieved with the 1-K stage and shield running at 3~K. The temperature of the bolometer wafer holder was 326~mK. The 8-mK gradient is due to the thermal resistance of the heat strap connecting the bolometer wafer holder to the cold head of the $^3$He refrigerator. With the 1-K stage and shield both at 1~K, we obtained a similar base temperature and a hold time longer than that of the 1-K EVR. 

The holder temperature achieved is sufficiently low compared to the transition temperature of the TES detectors to allow their nominal operation, as described in a companion paper  \cite{2020AA.QUBIC.PAPER4}. There, we also discuss the issue of microphonic noise induced by the pulse tubes. 
The hold time of the 0.3-K EVR at 318~mK exceeded 42~hours while carrying out the calibration of the QUBIC instrument. The time required to recycle the refrigerator is a few hours. 
This means that this cryogenic system fully supports night-time observations, with recycling during the day when scientific observations might be hampered by systematic effects due to the Sun in the sidelobes and/or thermal gradients on the forebaffle and window. If needed, the cryogenic system can even withstand 2 consecutive nights of observations without recycling. 

\subsection{Tilt tests}

The QUBIC polarimeter will scan the sky in azimuth, with different elevations in the range 20$^\circ-90^\circ$. Since the optical axis of the receiver is parallel to the axis of the cryostat, we tested the operation of the cryogenic system for tilts with respect to the vertical between 0$^\circ$-70$^\circ$. The temperature variation of the PT1 stage 2 is less than 50~mK for the entire required elevation range. The temperature of the 0.3-K stage changed by less than 5~mK. Both values are compatible with the requirements for the operation of the instrument. 

\section{Conclusions}

The cryogenic system of QUBIC described here implements a number of solutions that are of general interest for current CMB research, which involves large throughput detector arrays at sub-K temperatures. In particular, the operation without cryogenic liquids and with negligible vibrations, the large throughput of the optical path, the presence of a very heavy instrument to be cooled to $\sim$K and sub-K temperatures represent significant cryogenic challenges. The solutions adopted for the QUBIC cryostat allowed, for the first time, a successful demonstration of bolometric interferometry for CMB research. The system worked for many months in the laboratory, fulfilling all the requirements. The performance of the cryogenic system is satisfactory: the detectors run at 0.33~K, the polarization modulator runs at $\sim$10~K, the hold time of the 0.3-K EVR is $>$40~hours, with a recycling time of a few hours, allowing for continuous night-time observations once at the site. The next step will be the shipment of the TD to the the high mountain observation site of Alto Chorrillo. 

After on-sky validation in the TD configuration, the FI configuration of QUBIC will be implemented. There are two main differences between the TD and the FI from a cryogenic perspective. In the FI, the optical beam will have a larger throughput than in the TD, and the focal plane will host more detectors. 

As is evident from table \ref{tab:filters}, the size of the 300-K window in the TD is also the final (FI) one. For this reason the radiative heat load on the 40-K stage is already the final one. For the radiative heat load on the 3-K stage, we have to consider that the optical aperture area at 40~K will increase by a factor $\sim$3.3. Inspection of table \ref{tab:loadsmain} shows that, if we are able to keep the center of the 40-K filters cold, radiation from 40-K in the optical path will remain a subdominant contribution (3~mW will become 10~mW, still small with respect to the 80~mW we have from the rest of the 40-K shield).

The larger number of detectors (by a factor of 4) will not increase the load on the 1-K cooler, which is dominated by the CFRP tubes, nor on the 0.3-K cooler, which is dominated by stainless steel and CFRP tubes. 

For these reasons we are confident that for the FI configuration this cryostat will achieve a performance similar to the one described here in the TD configuration.

\acknowledgments
QUBIC is funded by the following agencies. France: ANR (Agence Nationale de la
Recherche) 2012 and 2014, DIM-ACAV (Domaine d’Intérêt Majeur-Astronomie et Conditions d’Apparition de la Vie), CNRS/IN2P3 (Centre national de la recherche scientifique/Institut national de physique nucléaire et de physique des particules), CNRS/INSU (Centre national de la recherche scientifique/Institut national et al de sciences de l’univers). Italy: CNR/PNRA (Consiglio Nazionale delle Ricerche/Programma Nazionale Ricerche in
Antartide) until 2016, INFN (Istituto Nazionale di Fisica Nucleare) since 2017.  Argentina: MINCyT (Ministerio de Ciencia, Tecnología e Innovación), CNEA (Comisión Nacional de Energía Atómica), CONICET (Consejo Nacional de Investigaciones Científicas y Técnicas).
 
D. Burke and J.D. Murphy acknowledge funding from the Irish Research Council under the Government of Ireland Postgraduate Scholarship Scheme.  D. Gayer and S. Scully acknowledge funding from the National University of Ireland, Maynooth. D. Bennett acknowledges funding from Science Foundation Ireland.

\bibliographystyle{JHEP}
\bibliography{qubic}

\begin{thebibliography}{10}
\bibitem{cernox}
``Cernox$^{TM}$ is a trademark of lakeshore cryotronic inc.''
\end{thebibliography}

\section*{APPENDIX A: Cryogenic system control electronics}

\subsection*{Temperature sensors}

Several temperature sensors are needed to monitor the temperature of the different parts of the cryogenic system of QUBIC. For temperatures above 3~K, we used Si diodes, which feature a wider measurement range with respect to Ge or Si thermistors, at the cost of a reduced sensitivity. The selected sensors are Si PNP transistors (BC856, only one junction used). These have been soldered to a 4-pin connector and encapsulated in a custom copper housing, with a through-hole for a 3-mm screw, used to firmly clamp the sensor to the measured part while minimizing thermal impedance. 16 diodes have been calibrated individually against a calibrated DT670 thermometer in a laboratory cryostat, and mounted as detailed in the first block of table \ref{tab:thermometers}. Their accuracy is in the 0.1~K range. 
\begin{table}[htbp]
\centering
\caption{\label{tab:thermometers} Type and location of the thermometers used in the QUBIC cryostat.}
\smallskip
\begin{tabular}{|r|l|l|l|}
\hline
\# & Location & Function & Type \\
\hline
T1 & 40-K filter & Cooldown monitoring & diode \\
T2 & 40-K shield (bottom) & Cooldown monitoring &  diode \\
T3 & PT1 stage 1 & Cooldown monitoring &  diode \\
T4 & PT2 stage 1 & Cooldown monitoring &  diode \\
T5 & 40-K shield (side) & Cooldown monitoring & diode \\
T6 & 3-K filter & Cooldown monitoring & diode \\
T7 & HWP 1 & Cooldown monitoring &  diode \\
T8 & HWP 2 & Cooldown monitoring &  diode \\
T9 & 3-K shield (down) & Cooldown monitoring & diode \\
T10 & 3-K shield (right) & Cooldown monitoring & diode \\
T11 & PT1 stage 2 & Cooldown monitoring &  diode \\
T12 & PT2 stage 2 & Cooldown monitoring &  diode \\
T13 & PT1 stage 2 & Sensitive T monitoring &  Cernox \\
T14 & PT2 stage 2 & Sensitive T monitoring &  Cernox \\
\hline
T15 & Heat switch 1-K - 3-K & Heat Switch 3 control & diode \\
T16 & Heat switch 1-K - 3-K & Heat Switch 4 control &  diode \\
\hline
T17 & Horns front & Sensitive T monitoring &  Ge NTD B \\
T18 & Horns back & Sensitive T monitoring &  Ge NTD B \\
T19 & RF switch block & T monitoring &  Ge NTD B \\
T20 & RF switch block  & T monitoring &  Ge NTD B \\
\hline
T21 & 1-K stage & Cooldown monitoring &  Cernox \\
T22 & 300-mK stage & Cooldown monitoring &  Cernox \\ 
T23 & Mirror M1  & Sensitive T monitoring &  Cernox \\
T24 & Mirror M2  & Sensitive T monitoring &  Cernox \\
\hline
T25 & $^4$He switch & 1-K $^4$He Refrigerator Control & diode \\
T26 & $^4$He cryopump & 1-K $^4$He Refrigerator Control &  diode \\
T27 & $^4$He cold head & 1-K $^4$He Refrigerator Control & Cernox \\
\hline
T28 & $^4$He cryopump & 0.3-K Refrigerator Control & diode \\
T29 & $^4$He heat switch & 0.3-K Refrigerator Control & diode \\
T30 & $^3$He cryopump & 0.3-K Refrigerator Control & diode \\
T31 & $^3$He heat switch & 0.3-K Refrigerator Control & diode \\
T32 & Film breaker & 0.3-K Refrigerator Control & RuO$_2$ \\
T33 & Cold Head & 0.3-K Refrigerator Control & Ge NTD E \\
\hline
\end{tabular}
\end{table}
Additional diodes are required to operate the gas-gap switches used to connect the 1-K stage to the 3-K stage (second block in table  \ref{tab:thermometers}). 

More sensitive thermometers (Cernox\cite{cernox} and Ge NTD\cite{1994SPIE.2198..630H}) are used to monitor the temperature of critical parts of the instrument (PT heads, Feedhorns array and RF switches, third block of table \ref{tab:thermometers}, and beam combiner mirrors M1 and M2, 4$^{th}$ block), the 1K (5$^{th}$ block) and the sub-K (6$^{th}$ block) evaporation refrigerators. At 4K, the Ge NTD B thermistor provides an accuracy of 0.2~mK and a sensitivity of 40~$\mu$K/$\sqrt{Hz}$. At 350~mK, the Ge NTD E thermistor provides an accuracy of 0.4~mK and a sensitivity of 70~nK/$\sqrt{Hz}$. 

\subsection*{Heaters}

Heaters are needed to operate the gas-gap heat switches connecting the 3-K and 1-K stages, the cryopump of the 1K refrigerator and its gas-gap heat switch, the cryopumps of the 0.3-K refrigerator and their gas-gap heat switches. They are listed in table \ref{tab:heaters}. The 330~$\Omega$ and 1~k$\Omega$ heaters are thick film power resistors, while the other heaters are phosphor bronze wire wound. The current for the heaters must run through a relatively low resistivity wire cable, to avoid dissipation in the cable itself. We selected 100~$\mu$m diameter Beryllium-Copper twisted pairs for the cable, as the best tradeoff between dissipated power and thermal conductivity. 

\begin{table}[htbp]
\centering
\caption{\label{tab:heaters} Type and location of the heaters used in the QUBIC cryostat.}
\smallskip
\begin{tabular}{|r|l|l|l|}
\hline
\# & Location & Function & Resistance \\
\hline
H1 & Heat switch 1-K - 3-K & Heat Switch 3 control & 330~$\Omega$ \\
H2 & Heat switch 1-K - 3-K & Heat Switch 4 control &  330~$\Omega$ \\
\hline
H3 & Horns & Thermal regulation  &  1~k$\Omega$ \\
\hline
H4 & $^4$He switch & 1-K Refrigerator Control & 330~$\Omega$  \\
H5 & $^4$He cryopump & 1-K Refrigerator Control &  330~$\Omega$  \\
\hline
H6 & $^4$He cryopump & 0.3-K Refrigerator Control & 400~$\Omega$  \\
H7 & $^4$He heat switch & 0.3-K Refrigerator Control & 2.2~k$\Omega$ \\
H8 & $^3$He cryopump & 0.3-K Refrigerator Control & 400~$\Omega$ \\
H9 & $^3$He heat switch & 0.3-K Refrigerator Control & 2.2~k$\Omega$ \\
\hline
\end{tabular}
\end{table}

\subsection*{Temperature sensors readout and heaters control}

The cooldown monitoring diodes feature an impedance of the order of 100~k$\Omega$. The manganin wires used to supply and read the diodes are several meters long, and feature resistances of the order of 1~k$\Omega$. Four-wires readout is thus mandatory. The custom readout electronics biases each diode independently, with a constant current of 10~$\mu$A (as required from the calibration) through 2 wires, and reads the voltage across the diode through other 2 wires. All the readout voltages are digitized using 18 bit ADCs, and output on a RS-232 interface. Since these temperatures are not expected to change fast, the readout voltages are output with a rate of about 1 data set per second. An AVS47 (PICOWATT) AC resistance bridge measures the Cernox and Ge sensors. The heaters are controlled by laboratory power supplies, interfaced, as the readout electronics, to a controller PC running the QUBIC studio control software \cite{2020AA.QUBIC.PAPER3}.

\subsection*{Harnesses}

\subsubsection*{Cryostat control harness}

The diode thermometers are biased and read by means of 75~$\mu m$ manganin wire cables assembled in looms. These produce negligible heat loads on the 1-K, 3-K and 40-K cold stages. The more sensitive Ge thermistors must be biased with a very low power AC, producing very small (in the $\mu V$ range) AC voltage signal. They are biased and read by means of 127~$\mu$m diameter phosphor-bronze double twisted pairs. The estimated length for all the thermometer harnesses is a few m, and efficient heat sinks on the different temperature stages have been provided. The heaters are connected to the outer world via 75~$\mu m$ manganin wire looms for the 2.2~k$\Omega$ and 1~k$\Omega$ heaters, while the lower resistance heaters have been connected via Be-Cu wires, 125~$\mu$m in diameter. All these cables are from 2 to 4 meters long. The resulting heat loads on the 40-K, 3-K, 1-K and 0.3-K stages is negligible, when compared to the heat loads of the detectors readout and aperture switches harnesses described below.

\subsubsection*{Detectors readout harness}

TES Detectors are read by means of a time-domain multiplexing electronics, as described in \cite{2020AA.QUBIC.PAPER4}. Each readout module reads 128 TESs, operating at 0.3~K. The connection between the detectors wafer on the 0.3-K stage and the 128 SQUIDs on the 1-K stage is provided by a 256 wires NbTi woven ribbon cable, 0.235~m long. For the full instrument (2048 detectors) 16 of these ribbons are needed, producing a total heat load of 6~$\mu$W on the $^3$He refrigerator. 

The connection between the SQUIDs at 1~K and the heat sink and breakout box at 3~K is provided again by NbTi woven ribbons, 0.5-m long for a total of 752 wires, and a total heat load of 10~$\mu$W if the heatsink on the 3-K stage is at 4~K. 

The connection between the heatsinks and breakout boxes at 3~K and the multiplexer ASICs on the 40-K stage is provided by 528 AWG36, 1~m long, phosphor-bronze twisted pairs, with 16 1~mm OD, 0.8~mm ID stainless steel tubes for shielding (total heat load 0.76~mW on the 3K stage); 128 AWG32, 1-m long phosphor bronze wires, with 64 stainless steel shields   (total heat load of 2.4~mW on the 3-K stage); 96 AWG32, 1-m long, phosphor bronze wires, with 48 AWG44, 1.5~mm OD, 1.4~mm ID phosphor bronze shields (total heat load of 4.1~mW on the 3-K stage).

The connection between the ASICs at 40K and the room-temperature feedthroughs is made of 240 AWG36, 1-m long phosphor bronze wires for the slow control, with 16 stainless steel shields, 0.8~mm OD, 0.6~mm ID, for a total heat load of 47~mW; 64 stainless steel coaxial cables for digital signals, 1-m long, 203~$\mu$m diameter, with 0.7~mm OD and 0.4~mm ID shield, for a total load of 64~mW; 32 AWG34 copper twisted pairs, 1-m long, with 32 stainless steel shields, 0.8~mm OD, 0.6~mm ID, for a total heat load of 185~mW; 48 AWG36 phosphor bronze twisted pairs, 1-m long, with 48 stainless steel shields, 1.5~mm OD, 14~mm ID, for a total heat load of 145~mW. 

\subsubsection*{Apertures switches harness}

The shutters for the apertures array are thermally connected to the 3-K stage. For 400 shutters (see \cite{2020AA.QUBIC.PAPER7}), 401 wires are necessary. The cable from 4~K to 40~K is made with AWG36 phosphor bronze wires, 1-m long, with 8 inox shields, 0.8~mm ID, 1~mm OD, producing a heat load on the 3-K stage of 11~mW. The cable from 40~K to 300~K is made in the same way, producing a heat load of 67~mW on the 40-K stage. 

\subsubsection*{Total heat loads from harnesses}

Summing the heat loads specified above, we obtain the following total heat loads from the harnesses: $\sim$6~$\mu$W on the 0.3-K stage ; $\sim$11~$\mu$W on the 1-K stage; $\sim$~11mW on the 3-K stage; $\sim$510~mW on the 40-K stage. In addition, the ASICs heat load on the 40-K shield is 0.4~W. All these loads have a minor impact on the heat load budget of each cryogenic stage. 

\end{document}